\documentclass[12pt]{iopart}

\pdfoutput=1

\usepackage[english]{babel}
\usepackage{
amssymb,amsfonts,latexsym}
\usepackage{graphicx}
\usepackage{color}
\usepackage{iopams}
\usepackage{subfigure}
\usepackage{afterpage}
\usepackage{blkarray}
\usepackage{cite}
\usepackage{tikz}
\usetikzlibrary{snakes}

\usepackage{times}
\usepackage[colorlinks,
citecolor=blue,linkcolor=red,urlcolor=blue]{hyperref}

\usepackage{cancel}

\newcommand{\be}{\begin{equation}}
\newcommand{\ee}{\end{equation}}
\newcommand{\bea}{\begin{eqnarray}}
\newcommand{\eea}{\end{eqnarray}}

\begin{document}

\title{Quasi-local conserved charges and spin transport in spin-$1$ integrable chains}
\author{Lorenzo Piroli and Eric Vernier}
\address{SISSA and INFN, via Bonomea 265, 34136 Trieste, Italy}
\ead{lpiroli@sissa.it, evernier@sissa.it}
\date{\today}

\begin{abstract}
We consider the integrable one-dimensional spin-$1$ chain defined by the Zamolodchikov-Fateev (ZF) Hamiltonian. The latter is parametrized, analogously to the XXZ spin-$1/2$ model, by a continuous anisotropy parameter and at the isotropic point coincides with the well-known spin-$1$ Babujian-Takhtajan Hamiltonian. Following a procedure recently developed for the XXZ model, we explicitly construct a continuous family of quasi-local conserved operators for the periodic spin-$1$ ZF chain. Our construction is valid for a dense set of commensurate values of the anisotropy parameter in the gapless regime where the isotropic point is excluded. Using the Mazur inequality, we show that, as for the XXZ model, these quasi-local charges are enough to prove that the high-temperature spin Drude weight is non-vanishing in the thermodynamic limit, thus establishing ballistic spin transport at high temperature.
\end{abstract}

\section{Introduction}
The importance of conserved quantities in the study of physical systems can hardly be overemphasized. When they are present, they provide additional constraints on the problem under investigation, sometimes resulting in the possibility of an exact solution. A prominent example is given by one-dimensional quantum integrable models solvable by the Bethe ansatz method, where an algebraic construction 
is known to provide an infinite discrete set of {\it local} conserved operators (or conserved charges) \cite{baxter, sutherland, gaudin, korepin, faddeev}.

As a result of the recent tremendous experimental advances in the field of ultra-cold atoms \cite{bloch, cazalilla, polkovnikov}, integrable models are no longer considered only of pure mathematical interest as they can now be realized in cold atomic laboratories with a very good approximation. As a consequence, the past decade has witnessed a renewed theoretical effort to understand the physical implications of integrability \cite{polkovnikov, cr-10}. In particular, it was realized that conserved charges play a fundamental role in the absence of thermalization in isolated integrable systems, where the long time steady state cannot be described by a thermal canonical ensemble. Rather a generalized Gibbs ensemble (GGE), taking into account a complete set of integrals of motion, has been proposed \cite{rigol-07} and a huge amount of studies have been devoted to prove or verify its validity \cite{cazalilla-06, cc-07, barthel-08, eckstein, cdeo-08, rigol-09, iucci-09, fioretto-10,cramer,cassidy,foini,cef-11,cef-12,eef-12,ck-12,mossel,gramsch,dora, brandino,fagotti-13, fa_es-13, goldstein-13,gurarie,kormos-13,collura,kcc-14,mussardo,ce-13,fa-14,fagotti-14,rajabpour-14,fcec-14, sotiriadis, sotiriadis_II, dwbc-14,wdb-14,pozsgay-14,mazza,pozsgay,bucciantini,goldstein, bertini,kz-15,dpc-15,dmv-15, rajabpour-15, langen-15,alba-15} (see also the review articles \cite{polkovnikov, gogolin} and references therein). The importance of the notion of locality of the conserved operators involved in the GGE construction has been first clarified in \cite{cef-12, fagotti-13}. 

The physical relevance of local conserved charges was also realized before in connection with the problem of transport in one-dimensional quantum systems \cite{mazur, suzuki, czp-95, znp}. In this framework a natural problem is the computation of the well known Drude weight $D$ , which can be identified with the asymptotic value of the time-average of the current auto-correlation function \cite{znp}. A non-vanishing value of $D$ signals dissipationless, or ballistic (as opposed to diffusive) transport. An important achievement has been the realization that in integrable systems the Drude weight can be bounded from below using the Mazur inequality \cite{mazur, suzuki} involving once again the conserved operators. 

However, for Bethe ansatz integrable systems, the application of the Mazur inequality using the known local charges might lead to a trivial vanishing bound for the Drude weight and an important case where this happens is provided by the spin Drude weight in the XXZ spin-$1/2$ chain with vanishing external magnetic field \cite{znp}. In fact, it can be shown that the local conserved operators, being even under spin inversion, give a vanishing contribution to the Mazur bound due to symmetry reasons (the spin-current being odd under this transformation). 
Understanding whether the spin Drude weight is zero or not at finite temperature in the XXZ model has indeed  attracted a large body of literature in the past years \cite{ss-90, nma,zotos-99,ag,h-m-03,bfks-05,spa,steinigeweg-09,steinigeweg-11, steinigeweg, znidaric-11, kbm-12, khlh-13, steinigeweg-14}. In the gapless regime, excluding the isotropic point, numerical studies and Bethe ansatz calculations resulted in the general agreement that the finite temperature spin Drude weight is non zero \cite{hhb-07}. Recently this was independently rigorously established by the discovery by Prosen of {\it quasi-local} conserved operators \cite{prosen-11}. The latter have been subsequently studied in \cite{ip-12, pi-13, mpp-14, prosen, pereira, mpp-15, imp-15, idw-15, zmp-15, iqdb-15}, where the ideas and techniques of Ref. \cite{prosen-11} have been extended and generalized (see also the recent work \cite{doyon}). The quasi-local operators found in these studies cannot be written as a sum over the chain of local densities acting on a {\it finite} number of neighboring sites (as it is the case for the previously known local conserved charges), but still exhibit properties that are typical of local operators, as we will discuss in the following sections.

The works \cite{prosen-11, ip-12, pi-13} considered the XXZ chain with open boundary conditions where quasi-local charges have been first discovered. These constructions have then been generalized in \cite{prosen, pereira} for periodic boundary conditions, which are those of interest in our work. Crucially, these quasi-local conserved operators are not invariant under spin inversion and, combined with the Mazur inequality, yield a non-vanishing spin Drude weight. It is now important to mention that very recently a different set of quasi-local conserved operators has been constructed for the XXZ model in Refs. \cite{imp-15, idw-15, iqdb-15}). These charges are even under spin inversion as the previously known local ones. Accordingly they give a vanishing contribution to the Mazur bound and are not immediately relevant for transport problems. However it has been unambiguously shown in \cite{idw-15} that in general they need to be included in the GGE construction in order to obtain the correct results for the long-time limit of local correlation functions.

From these discussions it is clear that the discovery of quasi-local charges in the XXZ model is far from being just a mathematical achievement, its physical consequences being immediate and far-reaching. For completeness, we mention that a systematic study of additional conservation laws and their physical implications has been performed for free fermionic XY spin chains in Refs. \cite{fagotti-14, fagotti_bertini-15,fagotti_collura, fagotti-16}. Furthermore, quasi-local integrals of motions (with a different unrelated meaning of quasi-locality) have also been recently discussed for many-body localized systems \cite{sba-13, imbrie, hno-13, ros} and in quantum field theories \cite{dlsb-15, emp-15,cardy-15}.

All of the constructions discussed so far have been developed in the prototypical XXZ spin-$1/2$ chain, while a natural question concerns higher spin generalizations. Here we consider the Zamolodchikov-Fateev (ZF) chain \cite{FZ}, arguably the simplest {\it integrable} spin-$1$ model. Note that a straightforward generalization of the XXZ Hamiltonian, obtained by replacing spin-$1/2$ with spin-$1$ operators, results in a non-integrable model, even though its low energy physics is described by the integrable nonlinear $O(3)$ $\sigma$-model. Transport properties of the non-integrable spin-$1$ chain and of its low energy effective theory have been studied in \cite{sd-97, sd-98, fujimoto, konik, kz-04}. 

It is well known that the ZF model can be constructed by a fusion procedure starting from the XXZ spin-$1/2$ chain \cite{fusion}. Analogously to the latter, the ZF model is parametrized by a continuous anisotropy parameter and at the isotropic point coincides with the spin-$1$ Babujian-Takhtajan Hamiltonian \cite{takhtajan, babujian}. Following Refs. \cite{prosen, pereira}, in this work we explicitly construct quasi-local conserved charges which are not invariant under spin-inversion and we show that these establish a non-zero spin Drude weight at high temperature through the Mazur inequality. Our constructions are valid for periodic boundary conditions in the gapless regime of the ZF Hamiltonian and for a dense set of commensurate values of the anisotropy parameter, where the isotropic point is excluded. The behavior of the Mazur bound computed in this work suggests that the spin Drude weight in the ZF model is a non-monotonic function of the absolute value of the anisotropy parameter. 

The rest of this manuscript is organized as follows. In section \ref{spin_1} we introduce the ZF Hamiltonian and review its derivation through the fusion procedure. We also introduce the algebraic tools needed in the following. In section \ref{quasi-local} the explicit construction of quasi-local charges is presented, while section \ref{mazur} is devoted to the computation of the Mazur bound for the spin Drude weight. Finally, conclusions are presented in section \ref{conclusions}.

\section{The spin-$1$ Zamolodchikov-Fateev model}\label{spin_1}
\subsection{The Hamiltonian and the algebraic Bethe ansatz}\label{sec:model}
The Zamolodchikov-Fateev model \cite{FZ} is defined as a spin-$1$ chain of $N$ sites with periodic boundary conditions and Hamiltonian 
\bea
\fl H_N=\sum_{j=1}^{N}\Big\{\left[s^{x}_js^{x}_{j+1}+s^{y}_js^{y}_{j+1}+\cos\left(2\gamma\right) s^{z}_js^{z}_{j+1}\right]+2\left[(s^{x}_{j})^{2}+(s^{y}_{j})^{2}+\cos\left(2\gamma\right)(s^{z}_{j})^{2}\right]\nonumber\\
 -\sum_{a,b}A_{ab}(\gamma)s^{a}_js^{b}_js^{a}_{j+1}s^{b}_{j+1}\Big\},
\label{hamiltonian}
\eea
where the indexes $a,b$ in the second sum take the values $x$, $y$, $z$ and where the coefficients $A_{ab}$ are defined by $A_{ab}(\gamma)=A_{ba}(\gamma)$ and
\be
A_{xx} = A_{yy} =1,\, A_{zz} = \cos\left(2\gamma\right),\quad  A_{xy}=1,\, A_{xz}=A_{yz}=2\cos\gamma-1.
\label{coefficients}
\ee
The Hilbert space is $\mathcal{H}_N=h_{1}\otimes\ldots\otimes h_{N}$, where $h_j\simeq \mathbb{C}^3$ is the local Hilbert space associated with spin $j$. The spin-$1$ operators $s^{a}_j$ are given by the standard representation of the $SU(2)$ generators, explicitly
\be
\fl s^x=\frac{1}{\sqrt{2}}\left(\begin{array}{c c c}0&1&0\\1&0&1\\0&1&0\end{array}\right),\quad s^y=\frac{1}{\sqrt{2}}\left(\begin{array}{c c c}0&-i&0\\i&0&-i\\0&i&0\end{array}\right),\quad s^z=\left(\begin{array}{c c c}1&0&0\\0&0&0\\0&0&-1\end{array}\right).
\label{spin_op}
\ee
In the following, we also define the local spin-$1$ basis as
\be
|\Uparrow \rangle =\left(\begin{array}{c}1\\0\\0\end{array}\right),\quad |0 \rangle =\left(\begin{array}{c}0\\1\\0\end{array}\right), \quad |\Downarrow \rangle =\left(\begin{array}{c}0\\0\\1\end{array}\right).
\ee
Introducing the anisotropy parameter 
\be
\Delta=\cos\gamma,
\ee 
the Hamiltonian (\ref{hamiltonian}) is well defined and Hermitian for $\Delta\in \mathbb{R}$. As we will see in section \ref{sec_fusion}, the corresponding spectrum is invariant under the transformation $\Delta\to-\Delta$. The ZF model is gapless for $-1\leq \Delta \leq 1$ (namely $\gamma$ a real parameter) and gapped for $|\Delta|>1$ ($\gamma$ purely imaginary) \cite{sogo, kirillov}. Throughout this paper we will always consider the gapless phase, namely $\gamma\in \mathbb{R}$. We also introduce for further use the parameter
\be
q=e^{i\gamma},
\label{q_parameter}
\ee
so that $\Delta=(q+q^{-1})/2$. It is easy to see that the Hamiltonian (\ref{hamiltonian}) commutes with the total magnetization $s^{z}_{T}=\sum_{j}s^{z}_j$. For $\Delta=1$ the Hamiltonian becomes
\be
H_N({\Delta=1})=\sum_{j=1}^{N}\left[{\bf s}_j\cdot {\bf s}_{j+1}-\left({\bf s}_j\cdot {\bf s}_{j+1}\right)^2\right]+4N,
\label{babujian}
\ee
namely it coincides with the well known $SU(2)$-invariant Babujian-Takhtajan Hamiltonian \cite{takhtajan, babujian} (see also Ref. \cite{vc-14}, where the corresponding dynamical structure factors were recently computed).

The Hamiltonian (\ref{hamiltonian}) can be exactly diagonalized using Bethe ansatz and it is most conveniently studied using the algebraic Bethe ansatz (ABA) approach \cite{korepin, faddeev}, which also allows one to construct local conserved operators as we now briefly review (we refer to \cite{korepin,faddeev} for a more detailed treatment). The central object of the ABA construction is the $R$-matrix, an operator $\mathcal{R}_{ij}(\lambda)$ acting on the tensor product of two local Hilbert spaces $h_i\otimes h_j$ and depending on a spectral parameter $\lambda$. The $R$-matrix has to satisfy the Yang-Baxter equations
\be
\mathcal{R}_{12}(\lambda)\mathcal{R}_{13}(\lambda+\mu)\mathcal{R}_{23}(\mu)=\mathcal{R}_{23}(\mu)\mathcal{R}_{13}(\lambda+\mu)\mathcal{R}_{12}(\lambda).
\label{yang_baxter}
\ee
Given a $R$-matrix, following the prescription of ABA \cite{korepin, faddeev} one introduces an auxiliary Hilbert space $h_0$ (along with the physical Hilbert spaces $h_j$ associated with the local spins) and the {\it fundamental} $L$-operators $\mathcal{L}_{0j}(\lambda)=\mathcal{R}_{0j}(\lambda)$ acting on $h_0\otimes h_j$, $j=1,\ldots , N$. The so called transfer matrix $\tau(\lambda)$ is then obtained as the partial trace over the auxiliary space of the product of $L$-operators along the chain
\be
\tau(\lambda)=\tr_{0}\left\{\mathcal{L}_{0N}(\lambda)\ldots\mathcal{L}_{01}(\lambda)\right\}.
\label{fundamental_transfer}
\ee
Using the Yang-Baxter equations (\ref{yang_baxter}), one can show that $\{\tau(\lambda)\}_{\lambda}$ is a family of commuting operators, namely
\be
[\tau(\lambda),\tau(\mu)]=0.
\label{commuting_tau}
\ee 
The Hamiltonian $H$ of the model corresponding to the initial $R$-matrix is then obtained, up to additive or multiplicative constants, by the logarithmic derivative in the spectral parameter of the transfer matrix $\tau(\lambda)$
\be
H\sim \frac{d}{d\lambda}\log\tau(\lambda)\Big|_{\lambda=\lambda^*},
\ee
where $\lambda^*$ is a value which depends on the specific form of the $R$-matrix and which is chosen in such a way that the Hamiltonian $H$ is local. Finally, it can be shown \cite{korepin} that an infinite set of local conserved operators is obtained, up to multiplicative or additive constants, by higher derivatives 
\be
Q_k\sim\frac{d^{k}}{d\lambda^{k}}\log\tau(\lambda)\Big|_{\lambda=\lambda^*}.
\ee
Locality here means that $Q_k$ is a translationally invariant operator that can be written as the sum along the chain of some density acting on $k$ neighboring sites, namely $Q_k=\sum_{j}q_j^{(k)}$, with $q^{(k)}_j$ a $k$-site supported operator acting as the identity on the rest of the chain.

The $R$-matrix corresponding to the Hamiltonian (\ref{hamiltonian}), was first discovered by directly solving the Yang-Baxter equations (\ref{yang_baxter}) in \cite{FZ}. It is, however, naturally obtained using the so called fusion procedure \cite{fusion}, which is by now well known in the literature (see for example Ref. \cite{kns-13,hagendorf}). In the following we review the fusion procedure and explicitly exhibit the $R$-matrix of the ZF model.

\subsection{The fusion procedure}\label{sec_fusion}
The results reviewed in this section are naturally discussed within the representation theory of the quantum group $U_q(sl_2)$ \cite{saleur, gomez}. Here we present an elementary treatment, invoking the general theory only when necessary.

Our starting point is the $R$-matrix of the XXZ model which we write as 
\be
R^{\left(\frac{1}{2},\frac{1}{2}\right)}(\lambda)
=
\left(
\begin{array}{cccc}
[\lambda+1] & 0 & 0 & 0 \\
0& [\lambda] & 1 & 0 \\
0 & 1 &[\lambda] & 0 \\
0 & 0 & 0 & [\lambda+1] \\
\end{array}
\right) \,.
\label{eq:R1212}
\ee
In the above equation and throughout all this work we use the conventional notation for the $q$-deformed numbers 
\be
[x] = \frac{q^x - q^{-x}}{q-q^{-1}}=\frac{\sin(\gamma x)}{\sin(\gamma)} \,,
\label{q_group_notation}
\ee
where $q$ is defined in (\ref{q_parameter}). The $R$-matrix (\ref{eq:R1212}) acts on the tensor product $\mathbb{C}^2\otimes\mathbb{C}^2$. The idea of the fusion procedure is to construct iteratively a solution of the Yang-Baxter equation (\ref{yang_baxter}) $R^{\left(\frac{m}{2},\frac{n}{2}\right)}$ acting on $\mathbb{C}^{m+1}\otimes\mathbb{C}^{n+1}$, where the starting point is provided by $R^{\left(\frac{1}{2},\frac{1}{2}\right)}$ in (\ref{eq:R1212}). The $R$-matrix of the ZF model then will correspond to the operator $R^{(1,1)}$ acting on $\mathbb{C}^3\otimes\mathbb{C}^3$.

In the following, we introduce the graphical representation according to which the $R$-matrix (\ref{eq:R1212}) can be represented as \cite{baxter, hubbard}
\be
R^{\left(\frac{1}{2},\frac{1}{2}\right)}_{12}(\lambda) = 
\begin{tikzpicture}[baseline={([yshift=-.0ex]current bounding box.center)}]
\draw (0,0) node[left] {\tiny$1$} -- (1,0);
\draw  (0.5,-0.5) node[shift={(0,-0.2)}] {\tiny$2$}  -- (0.5,0.5); 
\node at (0.24,0.28) {\tiny $\lambda$};
\end{tikzpicture}
\ee
The tensor product of two local spin-$1/2$ Hilbert spaces, associated with the spins labelled by $\alpha$ and $\beta$, can be decomposed into the sum of an antisymmetric spin-$0$ representation and a symmetric spin-$1$ representation, symbolically $\left(\frac{1}{2}\right)_{\alpha}\otimes \left(\frac{1}{2}\right)_{\beta} = \left(0\right)_{\alpha \beta} \oplus \left(1\right)_{\alpha \beta}$. One then notices that $R_{\alpha \beta}^{\left(\frac{1}{2},\frac{1}{2}\right)}(-1) = -2 P^0_{\alpha \beta}$, where $P^0_{\alpha \beta}$ is the projector operator over the antisymmetric spin-$0$ representation $(0)_{\alpha \beta}$ . Now, the $R$-matrix $R^{\left(\frac{1}{2},\frac{1}{2}\right)}$ satisfies the Yang-Baxter equation 
\be
\fl R^{\left(\frac{1}{2},\frac{1}{2}\right)}_{12}(\lambda)R^{\left(\frac{1}{2},\frac{1}{2}\right)}_{13}(\lambda+\mu)R^{\left(\frac{1}{2},\frac{1}{2}\right)}_{23}(\mu)=R^{\left(\frac{1}{2},\frac{1}{2}\right)}_{23}(\mu)R^{\left(\frac{1}{2},\frac{1}{2}\right)}_{13}(\lambda+\mu)R^{\left(\frac{1}{2},\frac{1}{2}\right)}_{12}(\lambda) \,,
\label{YBER12}
\ee
and choosing $\mu=-1$ we see from (\ref{YBER12}) that the operator
\be
R^{\left(\frac{1}{2},\frac{1}{2}\right)}_{13}(\lambda-1)R^{\left(\frac{1}{2},\frac{1}{2}\right)}_{12}(\lambda)
\label{product}
\ee
leaves stable the symmetric spin-$1$ representation in the decomposition $\left(\frac{1}{2}\right)_{2}\otimes \left(\frac{1}{2}\right)_{3} = \left(0\right)_{23} \oplus \left(1\right)_{23}$. As a consequence, the operator in (\ref{product}) defines a $R$-matrix acting on the tensor product $\left(\frac{1}{2}\right)_{1}\otimes \left(1\right)_{23}\simeq \mathbb{C}^{2}\otimes \mathbb{C}^{3}$, which we represent graphically as 
\be 
R^{\left(\frac{1}{2},1\right)}_{1(23)}(\lambda) = 
\begin{tikzpicture}[baseline={([yshift=-.0ex]current bounding box.center)}]
\draw (0,0) node[left] {\tiny$1$} -- (1,0);
\draw[double, thick]  (0.5,-0.5) node[shift={(0,-0.2)}] {\tiny$(23)$}  -- (0.5,0.5); 
\end{tikzpicture}
 = 
\begin{tikzpicture}[baseline={([yshift=-.0ex]current bounding box.center)}]
\draw (-0.25,0.) node[left] {\tiny$1$} -- (1.25,0);
\fill[gray!25] (0.5,-0.5) circle (10pt and 4 pt);
\draw  (0.25,-0.5) node[shift={(0,-0.2)}] {\tiny$2$}  -- (0.25,0.5); 
\draw  (0.75,-0.5) node[shift={(0,-0.2)}] {\tiny$3$}  -- (0.75,0.5); 
\node at (0.1,0.2) {\tiny $\lambda$};
\node at (0.9,0.2) {\tiny $\lambda-1$};
\end{tikzpicture}
 = 
\begin{tikzpicture}[baseline={([yshift=-.0ex]current bounding box.center)}]
\draw (-0.25,0.) node[left] {\tiny$1$} -- (1.25,0);
\fill[gray!25] (0.5,-0.5) circle (10pt and 4 pt);
\fill[gray!25] (0.5,0.5) circle (10pt and 4 pt);
\draw  (0.25,-0.5) node[shift={(0,-0.2)}] {\tiny$2$}  -- (0.25,0.5); 
\draw  (0.75,-0.5) node[shift={(0,-0.2)}] {\tiny$3$}  -- (0.75,0.5); 
\node at (0.1,0.2) {\tiny $\lambda$};
\node at (0.9,0.2) {\tiny $\lambda-1$}; 
\end{tikzpicture}\,,
\ee 
where the shaded ellipses mean that the tensor product of the local Hilbert spaces $2$ and $3$ is projected over the symmetric spin-$1$ representation. It is easily seen, for example graphically, that $R^{\left(\frac{1}{2},1\right)}$ obeys a Yang-Baxter equation of the form
\be
\fl R^{\left(\frac{1}{2},\frac{1}{2}\right)}_{12}(\lambda)R^{\left(\frac{1}{2},1\right)}_{13}(\lambda+\mu)R^{\left(\frac{1}{2},1\right)}_{23}(\mu)=R^{\left(\frac{1}{2},1\right)}_{23}(\mu)R^{\left(\frac{1}{2},1\right)}_{13}(\lambda+\mu)R^{\left(\frac{1}{2},\frac{1}{2}\right)}_{12}(\lambda) \,.
\label{YBER123}
\ee
Repeating the same procedure one can prove that the operator $R^{\left(\frac{1}{2},1\right)}_{1 3}(\lambda) R^{\left(\frac{1}{2},1\right)}_{2 3}(\lambda +1)$ acting on the product $\left( \frac{1}{2}\right)_1 \otimes \left( \frac{1}{2}\right)_2 \otimes 1_3$ leaves stable the symmetric representation $(1)_{12}$. As before, this allows us to define the operator $R^{(1,1)}$ acting on the tensor product $\mathbb{C}^{3}\otimes \mathbb{C}^{3}$, which in graphical notation is given by
\be 
R^{\left(1,1\right)}_{(12)3}(\lambda) = 
\begin{tikzpicture}[baseline={([yshift=-.5ex]current bounding box.center)}]
\draw[double, thick] (0,0) -- (1,0);
\draw[double, thick]  (0.5,-0.5)  -- (0.5,0.5); 
\end{tikzpicture}
 = 
\begin{tikzpicture}[baseline={([yshift=-.5ex]current bounding box.center)}]
\fill[gray!25] (0.5,-0.75) circle (10pt and 4 pt);
\fill[gray!25] (-0.25,0.) circle (4pt and 10 pt);
\draw (-0.25,-0.25)  -- (1.25,-0.25);f
\draw (-0.25,0.25)  -- (1.25,0.25);
\draw  (0.25,-0.75)  -- (0.25,0.75); 
\draw  (0.75,-0.75)  -- (0.75,0.75); 
\node at (0.2,0.35) {\tiny $\lambda$};
\node at (0.9,0.35) {\tiny $\lambda-1$};
\node at (0,-0.15) {\tiny $\lambda+1$};
\node at (0.9,-0.15) {\tiny $\lambda$};
\node at (-0.7,0.25) {\tiny$1$};
\node at (-0.7,-0.25) {\tiny$2$};
\node at (0.5,-1) {\tiny$3$};
\end{tikzpicture}
= 
\begin{tikzpicture}[baseline={([yshift=-.5ex]current bounding box.center)}]
\fill[gray!25] (0.5,0.75) circle (10pt and 4 pt);
\fill[gray!25] (0.5,-0.75) circle (10pt and 4 pt);
\fill[gray!25] (-0.25,0.) circle (4pt and 10 pt);
\fill[gray!25] (1.25,0.) circle (4pt and 10 pt);
\draw (-0.25,-0.25)  -- (1.25,-0.25);
\draw (-0.25,0.25)  -- (1.25,0.25);
\draw  (0.25,-0.75)  -- (0.25,0.75); 
\draw  (0.75,-0.75)  -- (0.75,0.75); 
\node at (0.2,0.35) {\tiny $\lambda$};
\node at (0.9,0.35) {\tiny $\lambda-1$};
\node at (0,-0.15) {\tiny $\lambda+1$};
\node at (0.9,-0.15) {\tiny $\lambda$};
\node at (-0.7,0.25) {\tiny$1$};
\node at (-0.7,-0.25) {\tiny$2$};
\node at (0.5,-1) {\tiny$3$};
\end{tikzpicture}
\ee
In order to have a symmetric $R$-matrix of the spin-$1$ integrable chain, we then finally perform a local similarity transformation
\be
\mathcal{R}(\lambda)=  \frac{1}{[\lambda][\lambda+1]} (M \otimes M)\cdot R^{(1,1)}(\lambda) \cdot (M \otimes M)^{-1},
\label{gauge}
\ee
where we performed a global rescaling by the prefactor $([\lambda][\lambda+1])^{-1}$ for future convenience and where 
\be 
M = \left(
\begin{array}{c c c}1&0&0\\
 0& \frac{\sqrt{[2]}}{\sqrt{2}}&0\\
0&0&1
\end{array}\right).
\label{gauge_matrix}
\ee
The $R$-matrix in (\ref{gauge}) can be explicitly written down in the (ordered) basis
\be
\fl \mathcal{B}=\left\{|\Uparrow,\Uparrow\rangle , |\Uparrow,0\rangle,  |\Uparrow,\Downarrow\rangle,|0,\Uparrow\rangle , |0,0\rangle,  |0,\Downarrow\rangle, |\Downarrow,\Uparrow\rangle , |\Downarrow,0\rangle,  |\Downarrow,\Downarrow\rangle\right\}
\ee
where it reads 
\be
\fl \mathcal{R}(\lambda)
=  \left(
\begin{array}{c c c c c c c c c}
a(\lambda)&0&0&0&0&0&0&0&0\\
0 & b(\lambda) & 0&c(\lambda)&0&0&0&0&0\\
0 & 0& d(\lambda)& 0&e(\lambda)&0&g&0&0\\
0 & c(\lambda) & 0&b(\lambda)&0&0&0&0&0\\
0 & 0 & e(\lambda)&0&f(\lambda)&0&e(\lambda)&0&0\\
0 & 0 & 0&0&0&b(\lambda)&0&c(\lambda)&0\\
0 & 0 & g& 0 &e(\lambda)& 0& d(\lambda)&0&0\\
0&0&0&0&0&c(\lambda)&0&b(\lambda)&0\\
0&0&0&0&0&0&0&0&a(\lambda)
\end{array}
\right),
\label{eq:R11explicit}
\ee
where
\bea
\fl a(\lambda)=[\lambda+1][\lambda+2],\quad b(\lambda)=[\lambda][\lambda+1],\quad c(\lambda)=[2][\lambda+1],\quad d(\lambda)=[\lambda][\lambda-1],
\nonumber
 \\
\fl e(\lambda)= [2][\lambda], \quad  f(\lambda)=[\lambda][\lambda+1]+[2],\quad g=[2],
\eea
and where as usual we have employed the square brackets notation defined in (\ref{q_group_notation}). The operator (\ref{eq:R11explicit}) can be explicitly seen to satisfy the Yang-Baxter equation (\ref{yang_baxter}), and it corresponds to the $R$-matrix of the ZF model. In particular, using the $R$-matrix (\ref{eq:R11explicit}) one can construct the fundamental Lax operator $\mathcal{L}(\lambda)$ and the corresponding transfer matrix $\tau(\lambda)$ as explained in the previous section, cf. Eq. (\ref{fundamental_transfer}) . The Hamiltonian (\ref{hamiltonian}) is then obtained as
\be
H_N= N+\frac{\sin(2\gamma)}{\gamma}\frac{{\rm d}}{{\rm d}\lambda}\log \tau(\lambda)\Big|_{\lambda=0}.
\ee

It is now important to notice a symmetry of the $R$-matrix (\ref{eq:R11explicit}) under the transformation $\Delta\to-\Delta$, or equivalently
\be
\gamma\to \gamma+\pi.
\label{symmetry}
\ee
In fact, introducing for the moment the explicit dependence of $\mathcal{R}$ on $\gamma$, it is straightforward to verify
\be
\mathcal{R}_{ij}(\gamma,\lambda)=-\left(\mathcal{N}_i\otimes\mathcal{N}_j\right)\mathcal{R}_{ij}(\gamma+\pi,\widetilde{\lambda})\left(\mathcal{N}_i\otimes\mathcal{N}_j\right)^{-1},
\label{symmetry_delta}
\ee
where 
\be
\widetilde{\lambda}=\frac{\lambda}{1+\pi/\gamma}
\label{new_lambda}
\ee
and
\be
\mathcal{N}=
\left(
	\begin{array}{c c c}
		i & 0 & 0\\
		0 & 1 & 0\\
		0 & 0 & i
	\end{array}
\right).
\ee
From (\ref{symmetry_delta}), following the transfer matrix construction outlined in section \ref{sec:model}, it is now straightforward to obtain
\be
H_N(\gamma+\pi)=\left(\mathcal{N}_1\otimes\ldots \otimes\mathcal{N}_N\right)^{-1}H_N(\gamma)\left(\mathcal{N}_1\otimes\ldots \otimes\mathcal{N}_N\right),
\ee
namely the spectrum of the Hamiltonian (\ref{hamiltonian}) is invariant under $\Delta\to-\Delta$, as we mentioned in section \ref{sec:model}.

\section{Quasi-local conserved charges}\label{quasi-local}
So far we have reviewed the ABA construction of the ZF Hamiltonian and the local conserved charges. As in the XXZ spin-$1/2$ model, one can prove that the latter are all invariant under the spin-inversion transformation defined as
\be
s_j^{\pm}\to s_j^{\mp},\qquad s^z_j\to -s^{z}_j.
\label{inversion}
\ee
More explicitly, introducing the unitary operator 
\be
U=e^{-i\pi\sum_{j=1}^{N} s^{x}_j},
\ee
one can see that the transformation (\ref{inversion}) is equivalent to
\be
s^{\alpha}_{j}\to U^{\dagger}s^{\alpha}_{j}U.
\label{transform}
\ee
The invariance of all the known local conserved charges with respect to spin-inversion follows from
\be
U^{\dagger}\tau(\lambda)U=\tau(\lambda),
\label{invariance}
\ee
which can be proved analogously to the spin-$1/2$ case (see for example \cite{pereira}). To be more explicit, we can define the following matrix acting on the auxiliary space $h_0$
\be
W_0=\left(
\begin{array}{c c c}
	0&0&1\\
	0&1&0\\
	1&0&0
\end{array}
\right),
\ee
and verify directly that
\be
W_0\mathcal{R}_{0j}(\lambda)W_0^{-1}=U\mathcal{R}_{0j}(\lambda)U^{\dagger}=e^{-i\pi s^{x}_j}\mathcal{R}_{0j}e^{i\pi s_j^{x}},
\label{identity_1}
\ee 
where $\mathcal{R}_{0j}(\lambda)$ is given in (\ref{eq:R11explicit}). Using now the definition (\ref{fundamental_transfer}), we have
\be
\fl U\tau(\lambda)U^{\dagger}=\tr_{0}\left\{U\mathcal{L}_{0N}(\lambda)\ldots\mathcal{L}_{01}(\lambda)U^{\dagger}\right\}=\tr_{0}\left\{W_0\mathcal{L}_{0N}(\lambda)\ldots\mathcal{L}_{01}(\lambda)W_0^{-1}\right\}=\tau(\lambda),
\ee
where we have used the cyclic property of the trace. As we will see in section \ref{mazur}, Eq. (\ref{invariance}) results in the fact that the local conserved charges provide a vanishing contribution to the Mazur bound. 

In this section, we explicitly construct {\it quasi-local} conserved charges which are not invariant under the spin-inversion (\ref{inversion}). As we already pointed out, we consider only the gapless regime, namely we choose $\gamma$ to be a real parameter and $-1\leq\Delta\leq 1$. Moreover, because of the symmetry discussed in the previous section, we can consider only the case $0\leq\Delta\leq 1$. Finally, we restrict to the (dense) set of values of $\Delta$ that correspond to $q$ being a root of unity. More precisely
\be
q = \mathrm{e}^{i \pi \frac{l}{m}},\qquad m, l\in \mathbb{Z}^{+},\quad 1\leq l< \frac{m}{2},   
\label{rational_condition}
\ee
with $l$, $m$ coprime (we remind that $q$ is defined in (\ref{q_parameter})). The condition $1\leq l< m/2$ is equivalent to requiring $0\leq \Delta\leq 1$. Our derivation of quasi-local operators follows that of Refs. \cite{prosen,pereira} for the XXZ spin-$1/2$ model with periodic boundary conditions, where similar restrictions on the anisotropic parameter were assumed.

Following the recent literature \cite{prosen}, in this work we distinguish between the properties of pseudo-locality and quasi-locality. The former is identified with the notion of extensivity of the Hilbert-Schmidt (HS) norm. Explicitly, we say that an operator $\mathcal{Q}$ is pseudo-local if
\be
||\mathcal{Q}||^2_{\rm HS}=\langle\mathcal{Q}^{\dagger}\mathcal{Q}\rangle =\frac{1}{d_0^{N}}\tr_{\mathcal{H}}\left\{\mathcal{Q}^{\dagger}\mathcal{Q}\right\}\sim \alpha N,\qquad \alpha\neq 0,
\label{extensivity}
\ee
$d_0$ being the dimension of the local Hilbert space (in our case $d_0=3$), and where the trace is taken over the whole Hilbert space (with dimension $d_0^N$).  The notion of quasi-locality is stronger than pseudo-locality and it also requires a decaying HS norm of the local densities with increasing support. More precisely, consider an operator $\mathcal{Q}$ written as
\be
\mathcal{Q}=\sum_{r=1}^{N}\sum_{x=0}^{N-1}\mathcal{P}^{x}\left(q_r\otimes {\rm id}_{N-r}\right),
\label{def_quasi_locality}
\ee
where $q_r$ is supported over the sites $1, \ldots ,r$, while ${\rm id}_{N-r}$ is the identity operator over the sites $r+1,\ldots, N$. Here $\mathcal{P}^{x}$ is the translation operator defined by
\be
\mathcal{P}^{x}(\mathcal{O}_j)=\mathcal{O}_{j+x}.
\label{def:translation}
\ee
We say that $\mathcal{Q}$ is quasi-local if there exist two positive real numbers $\xi,\kappa>0$ s.t. 
\be
||q_r||_{\rm HS}^{2}\leq \kappa e^{-\xi r}\,.
\label{def:densities}
\ee
From the definitions above, it is easy to see that if $\mathcal{Q}$ is quasi-local it is also pseudo-local.

The idea behind the construction of quasi-local conserved operators is similar, in spirit, to the one of the local charges previously outlined, the crucial difference being the use of {\it non-fundamental} (i.e. different to the $R$-matrix of the model) Lax operators to obtain a new transfer matrix commuting with the standard one. In particular, the auxiliary space, which we now denote with $\mathcal{A}$, will not be in general isomorphic to the local Hilbert space associated with the physical spin. This idea was used in Refs. \cite{prosen, pereira} where these non-fundamental Lax operators were obtained from non-unitary representations of the quantum group $U_q(sl_2)$.

The $R$-matrix intertwining two arbitrary representations of $U_q(sl_2)$ can be obtained by the general theory of quantum groups \cite{gomez}, from which one could derive the Lax operator for the spin-$1$ representation with a generic auxiliary representation $\mathcal{A}$ (which is the one needed in our case). As an alternative and more elementary derivation, in the following section we directly obtain the non-fundamental Lax operator by the fusion procedure previously described.

\subsection{Non-fundamental Lax operators}\label{non-fundamental}

We first recall that the quantum group $U_q(sl_2)$ is defined in terms of the three generators $S_z, S_+, S_-$ (not to be confused with the spin operators $s_j^{\alpha}$ in (\ref{spin_op})), satisfying the commutation relations \cite{gomez, saleur, pereira, prosen}
\bea
[S_z, S_\pm] = \pm S_\pm ,\label{comm_1}
\\
\left[S_+,S_-\right] =\frac{q^{2 S_z}- q^{-2 S_z}}{q-q^{-1}} = [2 S_z] .\label{comm_2}
\eea 
Note that in the limit $q\to 1$ Eqs. (\ref{comm_1}), (\ref{comm_2}) recover the standard $SU(2)$ commutation relations. When $q$ is a root of unity, cf. Eq. (\ref{rational_condition}) the representation theory of $U_q(sl_2)$ is richer than in the general case and additional irreducible representations exist. As in \cite{prosen,pereira}, here we are interested in the so called highest weight nilpotent representation \cite{gomez}. The latter is defined on a linear space $\mathcal{A}_m$ with $\dim \mathcal{A}_m=m$ ($m$ is defined by $q$ as in (\ref{rational_condition})), generated by the basis vectors $\{|0\rangle, \ldots , |m-1\rangle\}$. The action of the quantum group generators is defined by
\bea 
S_z |r\rangle = (r+v) |r \rangle ,\label{action_1}\\
S_+ |r\rangle = -[r+2v]|r+1 \rangle ,\label{action_2}\\
S_- |r\rangle = [r] |r-1\rangle ,\label{action_3}
\eea 
where the notation (\ref{q_group_notation}) is employed as usual and where $v$ is an arbitrary complex parameter. Note in particular that
\be
S^{+}|m-1\rangle=S^{-}|0\rangle=0.
\ee

The $R$-matrix intertwining between the auxiliary representation $\mathcal{A}_m$ and the physical spin-$1/2$ representation is well known and defines the corresponding $L$-operator which can be written as
\be
L^{\left(1\over 2\right)} (\lambda) =  
\left(
\begin{array}{cc}
 [\lambda + \frac{1}{2} + S_z] & S_-\\
S_+ & [\lambda + \frac{1}{2} - S_z]
\end{array}
\right), 
\label{non_fund_1/2}
\ee
as it was directly exploited in Refs. \cite{pereira, prosen}. In the spin-$1$ case of interest for us, one can build the $L$-matrix acting on the tensor product $\mathcal{A}_m\otimes h_j$ (where $h_j\simeq \mathbb{C}^3$ is the physical local Hilbert space) by the fusion procedure reviewed in the last section. 
Denoting the auxiliary representation $\mathcal{A}_m$ by a thick horizontal line, we readily construct $L^{(1)}(\lambda)$ from the product $L^{\left(1\over 2\right)} (\lambda-1)L^{\left(1\over 2\right)} (\lambda)$ as 
\be 
L^{(1)}(\lambda)  = 
\begin{tikzpicture}[baseline={([yshift=-.5ex]current bounding box.center)}]
\draw[line width=3pt] (0,0) -- (1,0);
\draw[double, thick]  (0.5,-0.5)  -- (0.5,0.5); 
\end{tikzpicture}
= 
\begin{tikzpicture}[baseline={([yshift=-.5ex]current bounding box.center)}]
\fill[gray!25] (0.5,-0.75) circle (10pt and 4 pt);
\draw[line width=3pt] (-0.25,0) -- (1.25,0);
\draw  (0.25,-0.75)  -- (0.25,0.75); 
\draw  (0.75,-0.75)  -- (0.75,0.75); 
\node at (0.1,0.25) {\tiny $\lambda$};
\node at (0.9,0.25) {\tiny $\lambda-1$};
\end{tikzpicture}
= 
\begin{tikzpicture}[baseline={([yshift=-.5ex]current bounding box.center)}]
\fill[gray!25] (0.5,0.75) circle (10pt and 4 pt);
\fill[gray!25] (0.5,-0.75) circle (10pt and 4 pt);
\draw[line width=3pt] (-0.25,0) -- (1.25,0);
\draw  (0.25,-0.75)  -- (0.25,0.75); 
\draw  (0.75,-0.75)  -- (0.75,0.75); 
\node at (0.1,0.25) {\tiny $\lambda$};
\node at (0.9,0.25) {\tiny $\lambda-1$};
\end{tikzpicture}
\ee
where we have used the graphical notation introduced in the previous section. In order to be consistent with (\ref{gauge}), we consider the Lax operator $L(\lambda)$ obtained after a similarity transformation
\be
L(\lambda)=ML^{(1)}(\lambda)M^{-1},
\ee
where $M$ is defined in (\ref{gauge_matrix}). Analogously to the spin-$1/2$ case (\ref{non_fund_1/2}), in the local spin basis $\{|\Uparrow\rangle, |0\rangle,|\Downarrow\rangle\}$, one can then explicitly write down $L(\lambda)$ as a $3\times 3$ matrix $L_{ij}$ whose entries are operators acting on the auxiliary space $\mathcal{A}_m$. They are given by
\bea
L_{11}&=&[\lambda +1/2 + S_z] [\lambda -1/2+ S_z],\label{l_1}\\
L_{12}&=&[2]^{1\over 2}S_-[\lambda-1/2 + S_z]\label{l_2},\\
L_{13}&=&(S_-)^2,\label{l_3}\\
L_{21}&=&[2]^{1\over 2}S_+[\lambda+1/2+S_z],\label{l_4}\\
L_{22}&=&S_+ S_- +  [\lambda + 1/2 + S_z] [\lambda -1/2- S_z],\label{l_5}\\
L_{23}&=&[2]^{1\over 2}S_-[\lambda+1/2-S_z] ,\label{l_6}\\
L_{31}&=&(S_+)^2,\label{l_7}\\
L_{32}&=&[2]^{1\over 2}S_+[\lambda-1/2-S_z],\label{l_8}\\
L_{33}&=& [\lambda +1/2 - S_z] [\lambda -1/2- S_z]\label{l_9},
\eea
where as usual we have employed the notation (\ref{q_group_notation}). Introducing the $3\times 3$ Kronecker matrices acting on the spin-$1$ physical space
\be
\left(E^{ab}\right)_{cd} = \delta^a_c \delta^b_d,
\ee 
the Lax operators is then rewritten as
\be
L(\lambda) = \sum_{a,b=1}^{3} L_{ab}(\lambda) E^{ab} \,,
\label{lax_2}
\ee
where $L_{a,b}$ are defined in (\ref{l_1})-(\ref{l_9}). As a check, we explicitly verified that the Lax operator above satisfies the Yang-Baxter relations involving the $R$-matrix (\ref{eq:R11explicit}). More precisely, we have the following identity in pictorial form 
\be
\begin{tikzpicture}[baseline={([yshift=-.5ex]current bounding box.center)}]
\draw[line width=3pt] (0,0) -- (2,0);
\draw[double, thick, rounded corners=5pt] (1.5,-1.5)-- (0.5,-0.5)  -- (0.5,0.5); 
\draw[double, thick, rounded corners=5pt] (0.5,-1.5)-- (1.5,-0.5)  -- (1.5,0.5); 
\end{tikzpicture}
\qquad
=
\qquad
\begin{tikzpicture}[baseline={([yshift=-.5ex]current bounding box.center)}]
\draw[line width=3pt] (0,0) -- (2,0);
\draw[double, thick, rounded corners=5pt] (1.5,1.5)-- (0.5,0.5)  -- (0.5,-0.5); 
\draw[double, thick, rounded corners=5pt] (0.5,1.5)-- (1.5,0.5)  -- (1.5,-0.5); 
\end{tikzpicture}
\label{eq:YBEpictorial}
\ee
In complete analogy with the standard case, Eq. (\ref{fundamental_transfer}), one can define the new transfer matrix
\be
\tau_{\mathcal{A}}(\lambda)=\tr_{\mathcal{A}}\left\{L_{\mathcal{A}N}(\lambda)\ldots L_{\mathcal{A}1}(\lambda)\right\},
\label{new_transfer}
\ee
where we dropped the subscript $m$ in the notation $\mathcal{A}_{m}$ for clarity. By repeated use of the relation in (\ref{eq:YBEpictorial}) (the so called ``train argument'') it is straightforward to see
\be
\left[\tau(\mu),\tau_{\mathcal{A}}(\lambda)\right]=0.
\label{commutation}
\ee 
As a consequence, $\tau_{\mathcal{A}}(\lambda)$ is a generator of conserved charges. However, they will be in general non-local. In the next section we show how a continuous family of quasi-local charges is obtained from $\tau_{\mathcal{A}}(\lambda)$ along the lines of \cite{prosen, pereira}.
\ \\
\subsection{Quasi-local charges}\label{sec:main_result}
We now explicitly exhibit the family of quasi-local conserved charges obtained in this work. We stress again that our construction is valid in the gapless regime (where we restrict without loss of generality to $0\leq \Delta \leq 1$) and for $q$ of the form (\ref{rational_condition}). The final result reads
\bea
\fl \mathcal{Q}(\lambda)= \sum_{\left\{a_j\right\},\left\{b_j\right\}}\left(\tr_{\mathcal{A}_m}\left\{\frac{{\rm d}}{{\rm d}v} \left(\widetilde{L}_{a_N b_N}(\lambda)\ldots \widetilde{L}_{a_1 b_1}(\lambda)  \right)\Big|_{v=0} \right\}\prod_{j=1}^{N}E_{j}^{a_jb_j}\right)
\nonumber \\
\hspace{5cm}- \tr_{\mathcal{A}_m} \left\{ \frac{{\rm d}}{{\rm d}v}\left( \frac{1}{3} \sum_{a=1}^{3} \widetilde{L}_{aa}(\lambda) \right)^N\Big|_{v=0} \right\}
,
\label{final_result}
\eea
where the sum is over all the sequences $\{a_j\}_{j=1}^{N},\{b_j\}_{j=1}^{N}$ with $a_j,b_j=1,2,3$ and where
\be
\widetilde{L}_{ab}(\lambda) = \frac{1}{[\lambda+1/2][\lambda-1/2]}L_{ab}(\lambda),
\label{tilde_l}
\ee
the operators $L_{ab}$ being defined in Eqs. (\ref{l_1})-(\ref{l_9}). The derivative in Eq. (\ref{final_result}) is with respect to the parameter $v$ defined by the Eqs. (\ref{action_1}), (\ref{action_2}). Finally, the parameter $\lambda$ takes value in the following subset of the complex plane  
\be
\left| \gamma |{\rm Re}(\lambda)| - \frac{\pi}{2}\right| < \frac{\pi}{2m},
\ee
or more explicitly
\be
\left| |{\rm Re}(\lambda)| - \frac{m}{2l}\right| < \frac{1}{2l},
\label{condition}
\ee
where $m$ and $l$ are defined by (\ref{rational_condition}). For $\lambda$ outside the domain (\ref{condition}) the operators (\ref{final_result}) are still well-defined, they still commute with the Hamiltonian (\ref{hamiltonian}) but are no longer quasi-local.

Our strategy to derive the charges (\ref{final_result}) and prove their quasi-locality follows closely that of Ref. \cite{pereira, prosen}. Our starting point is the auxiliary transfer matrix (\ref{new_transfer}) introduced in section \ref{non-fundamental} which can be easily rewritten as
\bea
\tau_{\mathcal{A}}(\lambda)  &=& \sum_{\left\{a_j\},\{b_j\right\}}\tr_{\mathcal{A}}\left\{L_{a_N b_N}(\lambda)\ldots L_{a_1 b_1}(\lambda)\right\}\prod_{j=1}^{N}E_{j}^{a_jb_j}
\,,
\eea
where as before the sum is over all the sequences $\{a_j\}_{j=1}^{N},\{b_j\}_{j=1}^{N}$ with $a_j,b_j=1,2,3$. Here and in the following we drop for convenience the subscript $m$ in $\mathcal{A}_m$. Next, we define the following traceless operator
\bea
\fl \mathcal{I}(\lambda,v)=\frac{1}{\varepsilon(\lambda)^{N \over 2}}  \left[\tau_{\mathcal{A}}(\lambda) - \frac{1}{3^N}\tr\left\{ \tau_{\mathcal{A}}(\lambda) \right\} \right]
\nonumber \\
\fl
=
\frac{1}{\varepsilon(\lambda)^{N \over 2}}  
\left[
\sum_{\left\{a_j\},\{b_j\right\}}\tr_{\mathcal{A}}\left\{ 
L_{a_N b_N}\ldots L_{a_1 b_1}
\right\}\prod_{j=1}^{N}E_{j}^{a_jb_j}
-  
 \tr_{\mathcal{A}} \left\{ \left( \frac{1}{3} \sum_{a=1}^{3} L_{a,a} \right)^N \right\}
\right]
 \,,
 \label{eq:Idef}
\eea
where we have dropped the explicit $\lambda$-dependence of the operators $L_{ab}$ and where we have introduced for future convenience a global rescaling, with $\varepsilon(\lambda)$ defined as
\be
\varepsilon(\lambda) = \left([\lambda+1/2][\lambda-1/2]\right)^2 \,.
\label{eq:tauT1}
\ee   
Following \cite{pereira}, it is instructive to compute the Hilbert-Schmidt norm of the operator $\mathcal{I}$. We have 
\bea
\langle 
\mathcal{I}^\dagger \mathcal{I}
\rangle 
\equiv
\frac{1}{3^N}
\tr\mathcal{I}^\dagger \mathcal{I} 
= \frac{1}{|\varepsilon(\lambda)|^N} \left[ \tr_{\mathcal{A} \otimes \mathcal{A}}
\left(\mathcal{L}_{\mathcal{A}\mathcal{A}}\right)^N
-
\left|
\tr_{\mathcal{A}}
\left(\mathcal{L}_{\mathcal{A}}\right)^N
\right|^2
\right]
\,,
\label{eq:IIT1T2}
\eea
where we have introduced the matrices 
\bea
\mathcal{L}_{\mathcal{A}\mathcal{A}}=\frac{1}{3}\sum_{a,b=1}^{3}L^{\ast}_{ab}\otimes L_{ab}  \,,
\label{eq:TAAdef}
\\
\mathcal{L}_{\mathcal{A}}=\frac{1}{3}\sum_{a=1}^{3}L_{aa}  \,,
\label{eq:TAdef}
\eea
acting respectively on $\mathcal{A}\otimes \mathcal{A}$ and on $\mathcal{A}$.

So far we have taken $v$ to be a generic complex parameter. We now consider the particular value $v=0$. As in the XXZ spin-$1/2$ case, this value is special since for $v=0$ the state $|r=0\rangle$ gets decoupled from the others in the auxiliary representation, cf. Eqs. (\ref{l_1})-(\ref{l_9}). Furthermore, we claim that for $v=0$ the following properties hold: 
\begin{enumerate}
\item \label{prop_1} $\mathcal{L}_\mathcal{A}$ is a diagonal matrix in the auxiliary space. For all values of $\lambda$ in the domain (\ref{condition}), its eigenvalue with the largest absolute value is non-degenerate and equal to $\varepsilon^{1/2}(\lambda)$, where $\varepsilon(\lambda)$ is defined in (\ref{eq:tauT1}). The corresponding eigenvector is $|0\rangle_{\mathcal{A}}$.
\item \label{prop_2} The state  $|00\rangle \equiv |0\rangle \otimes |0\rangle$ is a right eigenvector of the non-symmetric matrix $\mathcal{L}_{\mathcal{A}\mathcal{A}}$, with eigenvalue $|\varepsilon(\lambda)|$. We write the corresponding left eigenvector as $ \langle 00_L|$, normalized such that $ \langle 00_L | 00\rangle = 1$. For the special value $\lambda = \frac{m }{2 l}$, it can be explicitly written as
\be
 \langle 00_L| = \sum_{r=0}^{m-1} \left( 1-\frac{r}{m}\right) \langle r |\otimes\langle r | \,.
\label{eq:0L}
\ee 
\item \label{prop_3} In the domain (\ref{condition}), the eigenvalue $|\varepsilon(\lambda)|$ of $\mathcal{L}_{\mathcal{A}\mathcal{A}}$ corresponding to $|00\rangle$ is non-degenerate. Furthermore, it is the one with the strictly largest absolute value.
\end{enumerate}
Properties (\ref{prop_1}) and (\ref{prop_2}) can be checked analytically. Property (\ref{prop_3}) is analogous to the condition discussed in the XXZ spin-$1/2$ chain \cite{prosen,pereira}, where a domain analogous to (\ref{condition}) has been found for the spectral parameter of the quasi-local conserved charges. This has been rigorously analyzed in the spin-$1/2$ case in \cite{pi-13} for open boundary conditions and in \cite{prosen} for the periodic chain. Here we have verified condition (\ref{prop_3}) numerically up to $m=10$, $1\leq l\leq (m-1)/2$, $m,l$ coprimes, and a significant sample of values of $\lambda$.

From property (\ref{prop_3}) it is not difficult to see that, up to exponentially vanishing terms in $N$, one has
\bea
\left(\mathcal{L}_{\mathcal{A}\mathcal{A}}\right)^N &\sim& |\varepsilon(\lambda)|^N | 00 \rangle \langle 00_L |,
\label{eq:T1n}
\\
\left(\mathcal{L}_{\mathcal{A}}\right)^N &\sim& \varepsilon(\lambda)^{N \over 2} | 0 \rangle\langle 0 | \,,
\label{eq:T2n}
\eea
(see for example appendix C of \cite{pereira} for an explicit proof). It follows that the Hilbert-Schmidt norm of $\mathcal{I}$ is exponentially vanishing in the limit $N \to \infty$, cf. Eq. (\ref{eq:IIT1T2}), so that $\mathcal{I}$ is not a quasi-local operator.

In analogy with \cite{prosen, pereira} we then obtain an operator with linearly growing norm by deriving $\mathcal{I}(\lambda,v)$ with respect to the parameter $v$. Explicitly, we define
\be
\mathcal{Q} (\lambda) 
= 
\left. \frac{\partial}{\partial v} \mathcal{I}(v,\lambda) \right|_{v=0} \,.
\label{definition_aux}
\ee 
First, we note that the definition (\ref{definition_aux}) is easily seen to coincide with (\ref{final_result}). For what we said in the previous section $\mathcal{Q}(\lambda)$ commutes with the transfer matrix (\ref{fundamental_transfer}) and thus with the Hamiltonian (\ref{hamiltonian}). Next, we show that its norm is extensive. We have
\bea 
 \fl \frac{\langle \mathcal{Q}^{\dagger} \mathcal{Q} \rangle}{N}=
\tr_{\mathcal{A}\otimes\mathcal{A}} \left\{ \left(\widetilde{\mathcal{L}}_{\mathcal{A}\mathcal{A}}\right)^{N-1}\widetilde{\mathcal{L}}_{\mathcal{A}\mathcal{A}}^{(1,1)}   \right\} +  \sum_{n=0}^{N-2} \tr_{\mathcal{A}\otimes \mathcal{A}} \left\{ \widetilde{\mathcal{L}}_{\mathcal{A}\mathcal{A}}^{(1,0)} \left(\widetilde{\mathcal{L}}_{\mathcal{A}\mathcal{A}}\right)^{n} \widetilde{\mathcal{L}}_{\mathcal{A}\mathcal{A}}^{(0,1)}\left(\widetilde{\mathcal{L}}_{\mathcal{A}\mathcal{A}}\right)^{N-2-n}   \right\} 
\nonumber \\
 - 
N~ \left| \tr_{\mathcal{A}} \left\{ \left(\widetilde{\mathcal{L}}_{\mathcal{A}}\right)^{N-1}  \widetilde{\mathcal{L}}_{\mathcal{A}}^{(1)}  \right\}  \right|^2
\,,
\label{temp}
\eea
where 
\bea
 \widetilde{\mathcal{L}}_{\mathcal{A}\mathcal{A}} = \frac{1}{|\varepsilon(\lambda)|} {\mathcal{L}}_{\mathcal{A}\mathcal{A}} 
 \,, \qquad 
  \widetilde{\mathcal{L}}_{\mathcal{A}} = \frac{1}{\varepsilon(\lambda)^{1 \over 2}} {\mathcal{L}}_{\mathcal{A}}  \,,
   \label{aux_1}\\
 \widetilde{\mathcal{L}}_{\mathcal{A}\mathcal{A}}^{(p,q)}=  
\left. \frac{\partial^p}{\partial v^p}
\frac{\partial^{q}}{\partial \bar{v}^{q}}
\widetilde{\mathcal{L}}_{\mathcal{A}\mathcal{A}}(v,\bar{v}) 
 \right|_{v=\bar{v} = 0} 
 \,,
 \qquad 
  \widetilde{\mathcal{L}}_{\mathcal{A}}^{(p)}=  
\left. \frac{\partial^p}{\partial v^p}
\widetilde{\mathcal{L}}_{\mathcal{A}}(v) 
 \right|_{v = 0}. \label{aux_2}
\eea
We now take the limit of large $N$. Applying (\ref{eq:T1n}), (\ref{eq:T2n}) and using the fact that $| 0 \rangle_{\mathcal{A}\mathcal{A}}$ is an eigenvector of both $ \widetilde{\mathcal{L}}_{\mathcal{A}\mathcal{A}}^{(1,0)} $ and $ \widetilde{\mathcal{L}}_{\mathcal{A}\mathcal{A}}^{(0,1)}$, cf. Eqs. (\ref{l_1})-(\ref{l_9}), we obtain 
\bea 
\fl \frac{\langle \mathcal{Q}^{\dagger} \mathcal{Q} \rangle}{N}
=
\langle 00_L| \widetilde{\mathcal{L}}_{\mathcal{A}\mathcal{A}}^{(1,1)} | 00\rangle
- \left|  \langle 0| \widetilde{\mathcal{L}}_{\mathcal{A}}^{(1)} | 0\rangle  \right|^2
\nonumber \\
 +  
 (N-1) \left[  \langle 00_L| \widetilde{\mathcal{L}}_{\mathcal{A}\mathcal{A}}^{(1,0)} |00\rangle\langle 00_L|\widetilde{\mathcal{L}}_{\mathcal{A}\mathcal{A}}^{(0,1)}| 00\rangle
 - \left|  \langle 0| \widetilde{\mathcal{L}}_{\mathcal{A}}^{(1)} | 0\rangle  \right|^2
  \right]
  + \ldots 
 \,,
\label{calculations}
\eea
where the dots indicate terms which are exponentially vanishing for large $N$. Using now the explicit expressions for the operators defined in (\ref{aux_1}) and (\ref{aux_2}) we see that the term proportional to $(N-1)$ in (\ref{calculations}) vanishes and the final result reads 
\bea
\fl \lim_{N \to \infty}
  \frac{\langle \mathcal{Q}^{\dagger}(\lambda) \mathcal{Q}(\lambda) \rangle}{N}
  = 
\langle 00_L| \widetilde{\mathcal{L}}_{\mathcal{A}\mathcal{A}}^{(1,1)} | 00\rangle
- \left|  \langle 0| \widetilde{\mathcal{L}}_{\mathcal{A}}^{(1)} | 0\rangle  \right|^2=\frac{2 \left(\frac{\pi l}{m}\right)^2}{9 \sin \left(\frac{\pi l}{m}\right)^2}
\nonumber \\
\fl \times
\frac{
\left(1 +  3|[2 \lambda]|^2 \right)
+
6|[2]|\left(|[\lambda+1/2]|^2 + |[\lambda-1/2]|^2 \right) \langle 00_L |1 1  \rangle 
+
6\langle 00_L | 22 \rangle 
}
{|[\lambda+1/2]  [\lambda-1/2]|^2}.
\label{norm_general}
\eea
Specializing to $\lambda = \frac{m}{2 l}$, for which $ \langle 00_L |$ is known, cf. (\ref{eq:0L}), we obtain the explicit expression 
\be
\lim_{N \to \infty}
  \frac{\langle \mathcal{Q}^{\dagger} \mathcal{Q} \rangle}{N}
= 
\frac{4 \left(\frac{\pi l}{m}\right)^2}{3 \sin \left(\frac{\pi l}{m}\right)^2}
\frac{
2[2][\frac{m-l}{2l}]^2(1-\frac{1}{m})
+
(1-\frac{2}{m})
+ \frac{1}{6}
}
{[\frac{m-l}{2l}]^4} \,.
\label{eq:normexact}
\ee
Eqs. (\ref{norm_general}), (\ref{eq:normexact}) establish the pseudo-locality of $\mathcal{Q}(\lambda)$ as defined in (\ref{extensivity}). The extensivity of the norm of $\mathcal{Q}(\lambda)$ obtained in (\ref{norm_general}) and (\ref{eq:normexact}) is a crucial requirement in order to have a non-vanishing Mazur bound for the spin Drude weight. This is discussed in the following section where we explicitly compute the Mazur bound using the charges (\ref{final_result}).

As we already announced, the charges (\ref{final_result}) not only are pseudo-local but also quasi-local, cf. the definition given by Eqs.~(\ref{def_quasi_locality}), (\ref{def:densities}). Since the physical relevance for high temperature transport of the charges (\ref{final_result}) follows simply by their linearly growing norm (i.e. pseudo-locality), the proof of their quasi-locality , which will not be used in the following, is reported in \ref{sec:quasi_locality_proof}. 

We stress again that (\ref{norm_general}) is valid only for $\lambda$ satisfying the condition (\ref{condition}). Moreover, the derivation of this section is valid for $0\leq \Delta<1$, i.e. $1\leq l <(m-1)/2$ (where $l$, $m$ are as in Eq. (\ref{rational_condition})). For $-1<\Delta<0$, namely $(m-1)/2< l< m$ one has a similar derivation, even though some care must be taken. Indeed, in this case the domain (\ref{condition}) has to be modified and one obtains quasi-local charges for
\be
\left| |{\rm Re}(\lambda)| - \frac{m}{l}\right| < \frac{1}{2l}, \qquad ({\rm if}\ (m-1)/2< l< m).
\label{condition2}
\ee
This is the domain where the eigenvalue with the largest absolute value of the matrices $\mathcal{L}_{\mathcal{A}\mathcal{A}}$ and $\mathcal{L}_{\mathcal{A}}$ coincides for $(m-1)/2< l< m$. Note that in this case the center of the strip in the complex plane described by (\ref{condition2}) corresponds to ${\rm Re}(\lambda)=m/l$ and not $m/2l$. Accordingly, for $-1<\Delta<0$ one can obtain the analytical expression for the HS norm analogously to (\ref{eq:normexact}) for the special value $\lambda=m/l$. 

\section{Mazur bound for the spin Drude weight}\label{mazur}
\label{sec:Mazur}

We now discuss the consequences of the existence of the charges (\ref{final_result}) on the transport properties of the ZF model. As in the XXZ spin-$1/2$ case, the properties of extensivity and the behavior under spin-inversion (\ref{inversion}) play a crucial role in the following discussion.

As we already pointed out, the Hamiltonian (\ref{hamiltonian}) commutes with the global magnetization $s^{z}_T=\sum_{j=1}^{N}s_j^{z}$ which is then conserved. However, the local spin density $s^{z}_j$ has a non-trivial time evolution determined by the Heisenberg equations of motion (setting $\hbar=1$)
\be
\fl \frac{{\rm d}}{{\rm d}t}s^{z}_{j}(t)=i\left[H,s^{z}_j\right]=i \left[\hat{h}_{j-1},s_j^z\right]+i \left[\hat{h}_j,s_j^z\right]=i \left[\hat{h}_{j-1},s_j^z\right]-i \left[\hat{h}_{j},s_{j+1}^z\right].
\ee 
Here $\hat{h}_j$ is the local density associated with the Hamiltonian (\ref{hamiltonian}), namely
\bea
\fl \hat{h}_j=\left[s^{x}_js^{x}_{j+1}+s^{y}_js^{y}_{j+1}+\cos\left(2\gamma\right) s^{z}_js^{z}_{j+1}\right]+2\left[(s^{x}_{j})^{2}+(s^{y}_{j})^{2}+\cos\left(2\gamma\right)(s^{z}_{j})^{2}\right]\nonumber\\
 -\sum_{a,b}A_{ab}(\gamma)s^{a}_js^{b}_js^{a}_{j+1}s^{b}_{j+1},
\label{local_hamiltonian}
\eea
where the coefficients $A_{ab}(\gamma)$ are defined in (\ref{coefficients}), so that $H=\sum_{j=1}^{N}\hat{h}_j$. The spin-current operator is obtained by a comparison with the continuity equation for the local spin density $s^{z}_j$ \cite{znp, steinigeweg-09}. Explicitly, we have
\be
\mathcal{J}=i\sum_{j=1}^N\left[\hat{h}_{j-1},s_{j}^{z}\right] \,,
\ee
After a straightforward calculation, the spin-current operator in the ZF model then reads
\be
\mathcal{J}= 
\mathcal{J}^{(1)} + \mathcal{J}^{(2)}+ \mathcal{J}^{(3)}  
=
\frac{i}{2}\sum_{\ell=1}^{N}\left(\mathcal{J}^{(1)}_{\ell-1,\ell}+\mathcal{J}^{(2)}_{\ell-1,\ell}+\mathcal{J}^{(3)}_{\ell-1,\ell}\right),
\label{total_current}
\ee
where
\bea
\fl \hspace{1cm} \mathcal{J}^{(1)}_{\ell-1,\ell}&=&s_{\ell-1}^{+}s_{\ell}^{-}-s_{\ell-1}^{-}s_{\ell}^{+} \label{current_1}
\\
\fl \hspace{1cm}\mathcal{J}^{(2)}_{\ell-1,\ell}&=&(s_{\ell-1}^-)^2(s_{\ell}^+)^2-(s_{\ell-1}^+)^2(s_{\ell}^-)^2 \label{current_2}\\
\fl \hspace{1cm} \mathcal{J}^{(3)}_{\ell-1,\ell} &=& (1-2\cos\gamma) \left\{(s^+_{\ell-1}s^{-}_{\ell}-s^-_{\ell-1}s^{+}_{\ell})s^{z}_{\ell-1}s^{z}_{\ell}+
s^{z}_{\ell-1}s^{z}_{\ell}(s^+_{\ell-1}s^{-}_{\ell}-s^-_{\ell-1}s^{+}_{\ell})\right\}. \label{current_3}
\eea
Here $s_j^{\pm}$ are defined by
\be
s^{\pm}_j=s^{x}_{j}\pm is^{y}_{j}.
\ee
Note that the operator $\mathcal{J}^{(1)}$ has the same form as the current operator in the XXZ spin-$1/2$ chain, while $\mathcal{J}^{(2)}$ and $\mathcal{J}^{(3)}$ are additional contributions arising from the commutation between $s^{z}_j$ and the quartic terms in (\ref{hamiltonian}). Also, note that $\mathcal{J}^{(1)}$, $\mathcal{J}^{(2)}$, $\mathcal{J}^{(3)}$ are all odd under the spin-inversion (\ref{inversion}). 

In the framework of transport in one-dimensional quantum systems, the spin Drude weight at finite temperature $T$ can be defined as the long-time average of the current-current correlation function \cite{znp}, namely
\be
D=\lim_{t\to\infty}\frac{1}{t}\int_{0}^{t}{\rm d}t' \frac{\langle \mathcal{J}(t')\mathcal{J}(0)\rangle_T}{2NT}.
\ee
Here $\langle\ldots\rangle_T$ indicates the thermal expectation value. Within linear response theory, a non-vanishing Drude weight signals that an induced current is prevented from decaying to zero at large times and in this case transport is said to be ballistic (as opposed to diffuse). In general, the computation of the Drude weight (or even establishing whether it is vanishing or not) represents a very difficult task \cite{nma,zotos-99,ag,h-m-03,bfks-05,spa,steinigeweg-09,steinigeweg-11, steinigeweg, znidaric-11, kbm-12, khlh-13, steinigeweg-14}. The Mazur inequality provides a way to establish a lower bound for $D$ in terms of a set of commuting conservation laws $\{\mathcal{Q}_k\}$ \cite{mazur, suzuki,znp}. Explicitly, it reads 
\be 
D \geq \frac{1}{2 N T}\sum_k \frac{\left|\left\langle  \mathcal{J} \mathcal{Q}_k \right \rangle_T \right|^2 }{\left\langle  \mathcal{Q}_k^\dagger \mathcal{Q}_k \right \rangle_T}  \,,
\label{mazur_bound}
\ee
where $\{\mathcal{Q}_k\}$ are a set of conserved operators that are orthogonal in the sense $\langle\mathcal{Q}_k^{\dagger}\mathcal{Q}_\ell\rangle_T=\delta_{k\ell}\langle\mathcal{Q}^{\dagger}_k\mathcal{Q}_k\rangle_T$.

As we discussed in section \ref{quasi-local}, in the ZF model the local conserved charges derived from the transfer matrix (\ref{fundamental_transfer}) are even under the transformation (\ref{inversion}). As a result, they are orthogonal to the odd spin-current operator and give a vanishing contribution to the Mazur bound (\ref{mazur_bound}). Instead, the quasi-local charges (\ref{final_result}) are not even under (\ref{inversion}) (as Eq. (\ref{identity_1}) does not hold when one chooses as auxiliary space the representation $\mathcal{A}_m$ with general $v$). Along the lines of Refs. \cite{prosen, pereira}, where the spin-$1/2$ case was considered, we now show that the quasi-local operators constructed here provide a non-vanishing Mazur bound which we explicitly compute at high-temperature. As a result, we will establish high-temperature ballistic transport in the ZF model for $-1<\Delta<1$, $\Delta\neq 0$.

At high temperature, the Mazur inequality (\ref{mazur_bound}) for a single conserved operator $\mathcal{Q}(\lambda)$ can be written up to smaller corrections in $O({1 \over T})$, as
 \be 
D \geq \frac{1}{2 NT} \frac{\left|\left\langle  \mathcal{J}\mathcal{Q}(\lambda) \right \rangle \right|^2 }{\left\langle  \mathcal{Q}(\lambda)^\dagger\mathcal{Q}(\lambda)  \right \rangle} \,,
\label{infinite_temp}
\ee
where $\langle\ldots \rangle$ represents the infinite temperature expectation value as in (\ref{extensivity}), namely
\be
\langle\mathcal{O}\rangle=\frac{1}{3^{N}}\tr\mathcal{O}.
\ee 

\begin{figure}
\centering
\includegraphics[scale=1.]{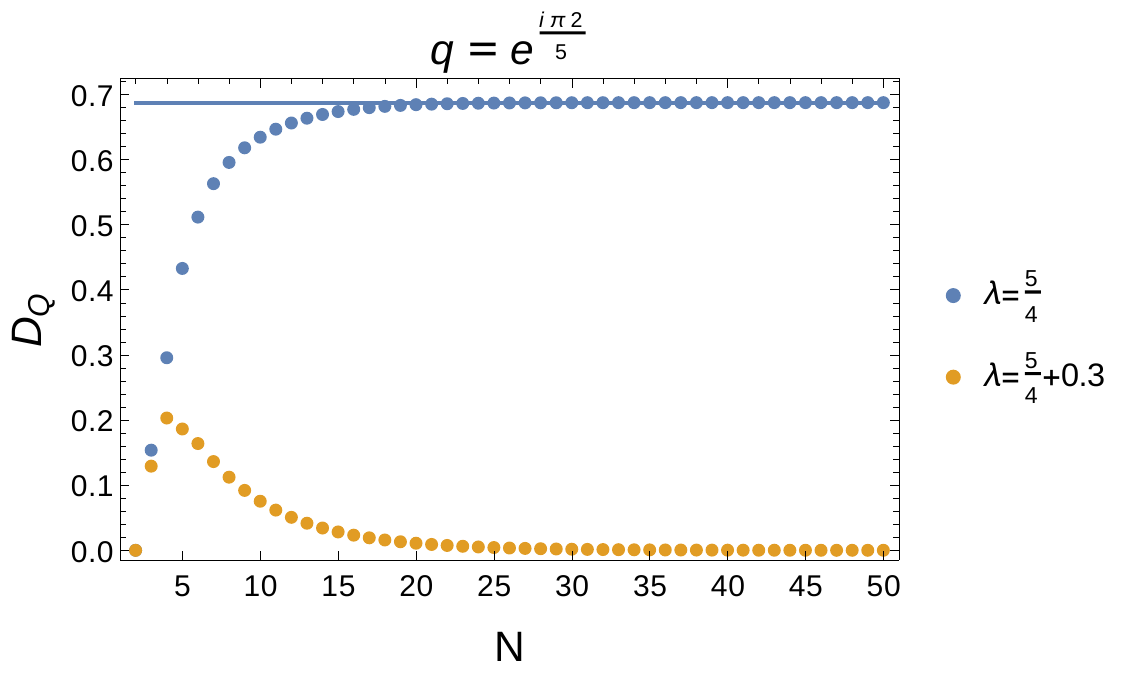}
  \caption{(Color online) Mazur bound (\ref{mazur_dq}) evaluated with a single quasi-local charge $Q(\lambda)$ for finite systems (where $N$ is the number of sites), for $\lambda = \frac{m}{2 l}$ (blue dots), and for a value of the spectral parameter outside the domain (\ref{condition}) (orange dots). The plot corresponds to $m=5$, $l=2$. Here the Mazur bound is computed numerically from Eqs. (\ref{temp}), (\ref{eq:J1I})-(\ref{eq:J3I}) and (\ref{eq:I1Jpart1part2}). We see that for large system sizes the Mazur bound for $\lambda = \frac{m}{2 l}$ approaches the analytical prediction (\ref{eq:DI1exact}) (blue solid line).}
 \label{fig:finite_size}
 \end{figure}

We now explicitly compute the r.h.s. of (\ref{infinite_temp}). From here on we set $\lambda=m/2l$ for which an explicit analytical expression for the Mazur bound can be obtained. Also, we will see that this choice maximizes the Mazur bound when only one quasi-local charge is considered. We need to compute
\be
\left\langle  \mathcal{J}\mathcal{\mathcal{Q}}(\lambda) \right \rangle  
=
\left. \frac{\partial}{\partial v}\left\langle  \mathcal{J}\mathcal{I}(v,\lambda)\right \rangle    \right|_{v=0} \,,
\label{eq:JI1JI}
\ee
where $\mathcal{J}$ is defined in (\ref{total_current}) and where $\mathcal{I}$ is given in (\ref{eq:Idef}). From Eqs. (\ref{current_1})-(\ref{current_3}) we see that each contribution $\mathcal{J}^{(k)}$ is traceless and therefore the identity term in $\mathcal{I}$ does not contribute to the scalar product $\left\langle \mathcal{J}^{(k)} \mathcal{I}\right\rangle $. We are then left with 
\bea
\fl \left\langle \mathcal{J}^{(k)} \mathcal{I}\right\rangle  
&=& \frac{1}{3^N }\frac{i}{2} \sum_{\ell=1}^N 
\frac{1}{\varepsilon(\lambda)^{N \over 2}}  \sum_{\left\{a_j\right\},\left\{b_j\right\}}\tr_{\mathcal{A}}\left\{  
L_{a_N b_N}\ldots L_{a_1 b_1}
\right\}
\tr\left\{ \mathcal{J}^{(k)}_{\ell-1,\ell} \prod_{j=1}^{N}E_{j}^{a_j b_j}\right\}
\nonumber
 \\
\fl
 &=& 
\frac{N}{3^2} 
\frac{i}{2}\frac{1}{\varepsilon(\lambda)^{N \over 2}} 
 \sum_{a_{1},b_{1},a_{2},b_{2}} 
\tr_{\mathcal{A}}\left\{L_{a_{2} b_{2}}L_{a_1 b_1}({ \mathcal{L}_{\mathcal{A}}})^{N-2} \right\}
\tr\left\{ \mathcal{J}^{(k)}_{1,2} E_{1}^{a_{1}b_{1}}E_{2}^{a_{2}b_{2}}\right\} \,,
\label{finite_size}
\eea
where $L_{ab}$ and $\mathcal{L}_{\mathcal{A}}$ are defined respectively in (\ref{l_1})-(\ref{l_9}) and (\ref{eq:TAdef}). Introducing
\bea 
{L_+} &=& \frac{1}{\sqrt{2}} \left(L_{12}+L_{23}\right),
\\
{L_-} &=& \frac{1}{\sqrt{2}} \left(L_{32}+L_{21}\right),
\eea 
we have
\bea 
\left\langle \mathcal{J}^{(1)} \mathcal{I}\right\rangle  = \frac{2i N}{3^2} \frac{1}{\varepsilon(\lambda)^{N \over 2}}
\tr_{\mathcal{A}} \left\{  \left({L_+} {L_-} - {L_-} {L_+}\right)({\mathcal{L}_\mathcal{A}})^{N-2}\right\}
 \label{eq:J1I}
 \\
\left\langle \mathcal{J}^{(2)} \mathcal{I}\right\rangle  = 
\frac{2 i N}{3^2} \frac{1}{\varepsilon(\lambda)^{N \over 2}}
\tr_{\mathcal{A}} \left\{  \left(L_{31} L_{13} - L_{13} L_{31}\right)({\mathcal{L}_\mathcal{A}})^{N-2} \right\}
 \label{eq:J2I}
\\
\left\langle \mathcal{J}^{(3)} \mathcal{I}\right\rangle  =
 \frac{2i N}{3^2}\frac{1}{\varepsilon(\lambda)^{N \over 2}} \frac{1-2 \cos \frac{\pi l}{m}}{2}\nonumber\\
 \times \tr_{\mathcal{A}} \left\{  \left(L_{32} L_{12}+L_{21} L_{23} -L_{12} L_{32}-L_{23} L_{21}\right)({\mathcal{L}_\mathcal{A}})^{N-2} 
 \right\}
   \,,
 \label{eq:J3I}
\eea
We now take the derivative with respect to $v$ according to (\ref{eq:JI1JI}). We then need to compute the general expression
\bea
\fl
\frac{{\rm d}}{{\rm d}v}\tr_{\mathcal{A}} \left\{\widetilde{L}_{ab} \widetilde{L}_{a'b'}(\widetilde{\mathcal{L}}_\mathcal{A})^{N-2}\right\}\Big|_{v=0}
= \tr_{\mathcal{A}} \left\{(\widetilde{L}'_{ab} \widetilde{L}_{a'b'}+\widetilde{L}_{ab} \widetilde{L}'_{a'b'}) (\widetilde{\mathcal{L}}_\mathcal{A})^{N-2} \right\}\nonumber
\\
+\sum_{n=1}^{N-2}
 \tr_{\mathcal{A}} \left\{ \widetilde{L}_{ab} \widetilde{L}_{a'b'} (\widetilde{\mathcal{L}}_\mathcal{A})^{n-1}\widetilde{\mathcal{L}}_{\mathcal{A}}^{(1)}(\widetilde{\mathcal{L}}_\mathcal{A})^{N-2-n} \right\},
\label{eq:I1Jpart1part2}
\eea
where $\widetilde{L}_{ab}$, $\widetilde{\mathcal{L}}_{\mathcal{A}}^{(1)}$ are defined respectively in (\ref{tilde_l}), (\ref{aux_2}) and where we used
\be
\widetilde{L}'_{a'b'}=\frac{{\rm d}}{{\rm d}v}\widetilde{L}_{a'b'}\Big|_{v=0}.
\ee

Consider now the sum in (\ref{eq:I1Jpart1part2}). First, note that comparing with Eqs. (\ref{eq:J1I}), (\ref{eq:J2I}), (\ref{eq:J3I}), we always have $a\neq b$, $a'\neq b'$. Next, from Eqs. (\ref{l_1})-(\ref{l_9}) we have that if $a\neq b$ then $L_{a,b}|_{v=0}|0\rangle=0$, while it is immediate to see that $|0\rangle$ is an eigenstate of $\widetilde{\mathcal{L}}_{\mathcal{A}}^{(1)}$. Finally, from Eq. (\ref{eq:T2n}) we note that, up to exponentially small corrections in the large $N$ limit, we can substitute the trace over $\mathcal{A}$ in the sum with the expectation value on $|0\rangle$. Putting everything together we conclude that the sum in (\ref{eq:I1Jpart1part2}) is exponentially vanishing for large $N$. Using once again (\ref{eq:T2n}) we arrive at
\bea
 \lim_{N\to \infty}
\frac{{\rm d}}{{\rm d}v}\tr_{\mathcal{A}} \left\{\widetilde{L}_{ab} \widetilde{L}_{a'b'}(\widetilde{\mathcal{L}}_\mathcal{A})^{N-2}\right\}\Big|_{v=0}
=\langle 0|\widetilde{L}'_{ab} \widetilde{L}_{a'b'}+\widetilde{L}_{ab} \widetilde{L}'_{a'b'}  |0\rangle ,
\eea
which results in the following explicit formulas
\bea 
\fl \lim_{N\to \infty}
\frac{\left\langle \mathcal{J}^{(1)} \mathcal{Q}\right\rangle }{N}=(-i)
\frac{2}{[\lambda+1/2][\lambda-1/2]}\frac{\frac{\pi l}{m}}{\sin \frac{\pi l}{m} }  \frac{[2]}{3^2}
\times
{ \left( [\lambda+1/2]+[\lambda-1/2]  \right)^2  \over [\lambda+1/2][\lambda-1/2] },
\label{eq:J1Iexact}\\
\fl \lim_{N\to \infty}
\frac{\left\langle \mathcal{J}^{(2)} \mathcal{Q}\right\rangle }{N}=(-i)
\frac{2}{([\lambda+1/2][\lambda-1/2])^2}\frac{\frac{\pi l}{m}}{\sin \frac{\pi l}{m} }  \frac{[2]}{3^2} \times 2
\label{eq:J2Iexact}, \\
\fl \lim_{N\to \infty}
\frac{\left\langle \mathcal{J}^{(3)} \mathcal{Q}\right\rangle }{N}=(-i)
\frac{2}{[\lambda+1/2][\lambda-1/2]}\frac{\frac{\pi l}{m}}{\sin \frac{\pi l}{m} }  \frac{[2]}{3^2} 
\times \left(4\cos\frac{\pi l}{m} -2\right). 
\label{eq:J3Iexact}
\eea

\begin{figure}
\centering
\includegraphics[scale=0.7]{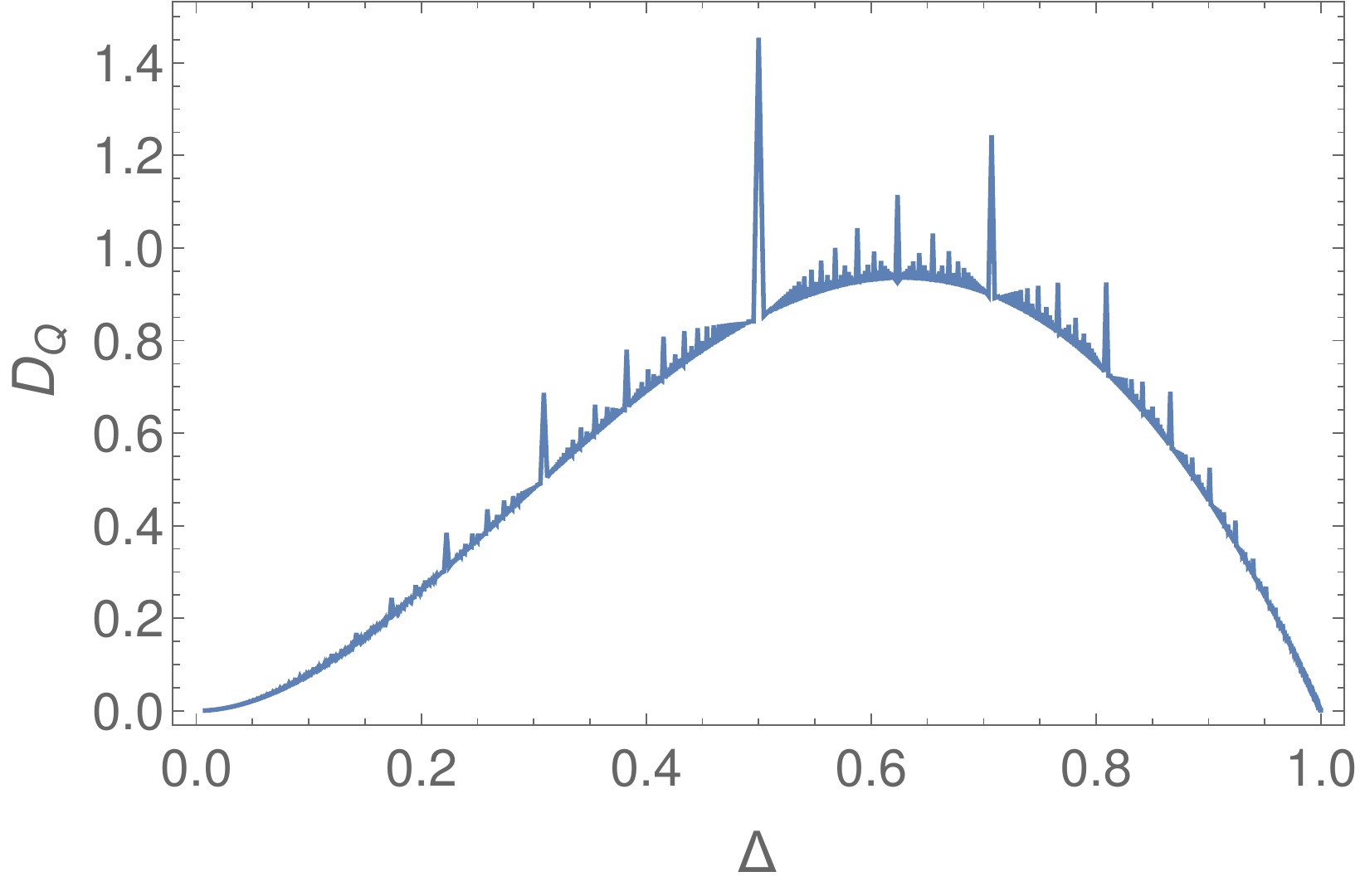}
  \caption{(Color online) Mazur bound  (\ref{eq:DI1exact}) in the thermodynamic limit, at the special value $\lambda = \frac{m}{2 l}$, plotted as a function of the anisotropy $\Delta = \cos\frac{\pi l}{m} $. We show only the regime $0 \leq \Delta <1$, the result being symmetric with respect to $\Delta = 0$. }
 \label{fig:Fractal}
 \end{figure}

We have now all the ingredients to write down the final result. First, we define
\be
D_{Q}(\lambda)= \frac{2}{N} \frac{\left|\left\langle  \mathcal{J}\mathcal{Q}(\lambda) \right \rangle \right|^2 }{\left\langle  \mathcal{Q}(\lambda)^\dagger\mathcal{Q}(\lambda)  \right \rangle},
\label{mazur_dq}
\ee
so that $D\geq D_{\mathcal{Q}}/4T$, cf. (\ref{infinite_temp}). Putting everything together, we arrive at an explicit analytical formula for the Mazur bound  in the thermodynamic limit, when the latter is computed with a single quasi-local charge with spectral parameter $\lambda=m/2l$. It reads
\bea
\fl 
\lim_{N\to\infty}D_{\mathcal{Q}}\left(\frac{m}{2l}\right)
=
\frac{64}{3}  \sin^2\left(\frac{\pi l}{2m} \right) \cos^2\left(\frac{\pi l}{m} \right)\nonumber\\
\times \left\{2\cos\left(\frac{\pi l}{m}\right)\left( 1 - \frac{1}{m}\right) + 2  \sin^2\left(\frac{\pi l}{2m} \right) \left(\left(  1 - \frac{2}{m}\right) + \frac{1}{6}\right) \right\}^{-1} \,. 
\label{eq:DI1exact}
\eea
We report in Fig. \ref{fig:finite_size} the comparison between the expression (\ref{eq:DI1exact}) and the Mazur bound for finite systems, as numerically computed from Eqs. (\ref{temp}), (\ref{eq:J1I})-(\ref{eq:J3I}) and (\ref{eq:I1Jpart1part2}),  at two different values of the spectral parameter. We see that for the special value $\lambda=m/2l$, within the domain (\ref{condition}), the numerical values for finite systems nicely approach the thermodynamic limit (\ref{eq:DI1exact}). On the other hand, for $\lambda$ outside the domain (\ref{condition}) the operator $\mathcal{Q}(\lambda)$ is no longer quasi-local and the Mazur bound approaches zero for $N\to\infty$. 

We remind here that the analytical expression (\ref{eq:DI1exact}) is valid for $0\leq \Delta <1$ for the commensurate values of the anisotropy corresponding to (\ref{rational_condition}). One can perform an analogous treatment in the case $-1<\Delta\leq 0$, namely $m/2<l<m$. As we discussed in section \ref{quasi-local}, the only difference would be the choice of the special point $\lambda=m/l$ instead of $\lambda=m/2l$ (see the discussion at the end of section \ref{quasi-local}). We explicitly checked that the Mazur bound obtained in this way is symmetric with respect to $\Delta=0$.

\begin{figure}
\centering
\includegraphics[scale=0.75]{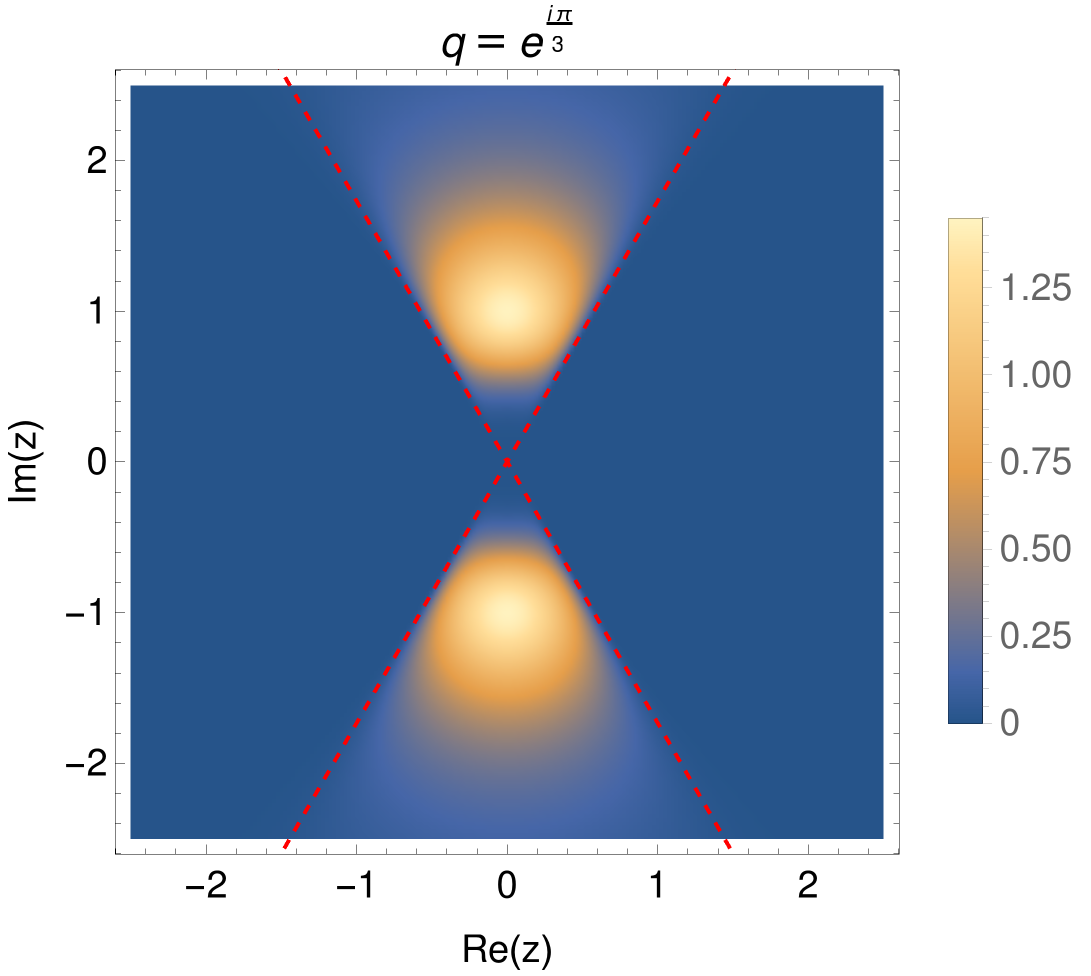}
\hspace{1cm}
  \caption{(Color online) Mazur bound (\ref{mazur_dq}) evaluated with a single quasi-local charge $Q(\lambda)$, plotted in the complex plane of the multiplicative spectral parameter $z = q^{\lambda}$, for $q = \mathrm{e}^{i \frac{\pi}{3}}$ (the plot corresponds to $\gamma=\pi/3$, namely $m=3$, $l=1$). Here the Mazur bound is computed numerically from Eqs. (\ref{temp}), (\ref{eq:J1I})-(\ref{eq:J3I}) and (\ref{eq:I1Jpart1part2}) for a system of $N=30$ sites. Blue regions correspond to a zero value of the Mazur bound, while the non-zero values are associated with brighter regions, located inside the cone $\left| |\arg z| - \frac{\pi}{2}\right| < \frac{\pi}{2m}$ (plotted in dashed red lines for comparison).}
 \label{fig:lightcone}
 \end{figure}

The result (\ref{eq:DI1exact}) is plotted as a function of $\Delta = \cos\frac{\pi l}{m}$ in Fig. \ref{fig:Fractal}, from which a fractal structure clearly emerges, analogously to the XXZ spin-$1/2$ chain. Note that in this section, the Mazur bound has been evaluated for a single conserved charge $\mathcal{Q}(\lambda)$, the final expression (\ref{eq:DI1exact}) being valid for $\lambda=m/2l$. However, a non-vanishing Mazur bound is in general obtained also for $\lambda$ satisfying condition (\ref{condition}), as we illustrated in Fig. \ref{fig:lightcone}. We have verified that the Drude weight with a single conserved operator is maximized by the choice $\lambda=m/2l$. This is once again analogous to the XXZ spin-$1/2$ chain \cite{pereira, prosen}. 

The continuous family of operators $\{\mathcal{Q}(\lambda)\}_{\lambda}$, with $\lambda$ satisfying (\ref{condition}), can be used to raise the bound (\ref{eq:DI1exact}) as done in \cite{pi-13, prosen}, where the sum in the r.h.s. of (\ref{mazur_bound}) was replaced by an integral over the spectral parameter. However, we claim that the improved Mazur bound would still be vanishing at $\Delta=0$. Indeed, we explicitly checked that the contribution of $\mathcal{Q}(\lambda)$ is vanishing for all $\lambda$ satisfying condition (\ref{condition}) at $\Delta=0$. Thus, an integral over the spectral parameter would still yield a vanishing Mazur bound at $\Delta=0$. In fact, as shown in Fig. \ref{fig:Fractal}, the behavior of the latter evaluated for a single charge seems to suggest that the spin Drude weight $D$ for the ZF model is non-monotonic as a function of the absolute value of the anisotropy parameter $\Delta$ (even though no definite conclusion on this point can be drawn, the Mazur inequality only providing a lower bound for $D$). Finally, note that the Mazur bound is also vanishing at the isotropic point $\Delta=1$, as for the spin-$1/2$ case.

\section{Conclusions}\label{conclusions}
In this work we have considered the simplest spin-$1$ integrable chain, the Zamolodchikov-Fateev model, which can be constructed by a fusion procedure starting from the XXZ spin-$1/2$ chain. In the gapless regime, for a dense set of values of the anisotropy parameter $\Delta$, we have shown that the techniques of Refs. \cite{prosen, pereira} can be applied to explicitly exhibit a family of quasi-local conserved charges. The latter are independent of the local ones obtained by the previously known algebraic Bethe ansatz methods.

The quasi-local charges are obtained by a transfer matrix construction, where the auxiliary space is a non-unitary representation of the quantum group $U_q(sl_2)$. As a result, the quasi-local charges are not even under the spin inversion (\ref{inversion}), thus providing a non-vanishing Drude weight through the Mazur bound inequality, in analogy with the XXZ spin-$1/2$ model \cite{prosen-11, pi-13, prosen, pereira}.

We have found that the Mazur bound establishes a non-vanishing Drude weight, and hence ballistic transport, in the range $-1<\Delta<1$, with $\Delta\neq 0$. Furthermore, the behavior of the Mazur bound, evaluated using the most relevant quasi-local conserved charge, seems to suggest that the Drude weight in the ZF model is a non-monotonic function of the absolute value of the anisotropy $\Delta$.

Recently, another type of conserved charges, which are even under the spin inversion (\ref{inversion}), have been discovered in the XXZ spin-$1/2$ model \cite{imp-15, idw-15, iqdb-15}. For these charges the independence with known local conserved operators does not easily follow from symmetry reasons and a more sophisticated technical treatment is required \cite{imp-15}. While they give a vanishing contribution to the Mazur bound, and are thus not immediately relevant for transport problems, they were shown to have direct impact on the relaxation dynamics of isolated integrable quantum systems \cite{idw-15, iqdb-15}, in particular for the GGE construction. A central piece of these recent investigations has been the computation of expectation values of these charges on the eigenstates of the system \cite{idw-15, iqdb-15}. We expect that these constructions can be formally generalized to the spin-$1$ case considered in our work. 

Finally, we mention that quasi-local conserved operators have also been discussed in the framework of quantum field theories \cite{dlsb-15, emp-15, cardy-15}. In particular, pseudo-local charges were studied in the free Klein-Gordon model in Ref.~\cite{dlsb-15}, where their implications on energy transport were investigated. In Ref.~\cite{emp-15} quasi-local conserved charges with a different notion of quasi-locality were explicitly obtained in operatorial form in the free Majorana fermion theory and their relevance for the GGE was analysed. It is still unclear whether quasi-local conserved operators similar to those discussed so far can be derived for interacting field theories and models defined on the continuum. This problem is particularly relevant for models displaying particle bound states, such as the sine-Gordon field theory or the attractive Lieb-Liniger gas \cite{ccaux-07,pozsgay, goldstein, piroli}. It would be extremely interesting to gain more insights into these issues in future studies. 

\section{Acknowledgments}
We would like to thank Ian Affleck, Pasquale Calabrese and Toma\v{z} Prosen for useful discussions and Michael Brockmann for comments on this manuscript. E.V. acknowledges support by the ERC under Starting Grant 279391 EDEQS

\appendix

\section{Proof of quasi-locality}\label{sec:quasi_locality_proof}
In section~\ref{quasi-local} we proved the pseudo-locality of the conserved charges (\ref{final_result}) derived in this work. As it is clear from our discussion in section~\ref{mazur}, this is enough to show their relevance for high temperature transport. However, the stronger property of quasi-locality, as defined by Eqs.~(\ref{def_quasi_locality}), (\ref{def:densities}) holds for the charges (\ref{final_result}). In this appendix we sketch the proof of this statement.

First, it is useful to define the following set of $3\times 3$ matrices acting on the physical spin-$1$ local space
\bea
\fl F^{\alpha\beta}=E^{\alpha\beta},\quad {\rm if\ }\alpha\neq \beta,\\
\fl F^{11}={\rm id}_{3},\qquad F^{22}= \left(
\begin{array}{c c c}1&0&0\\
 0& 0&0\\
0&0&-1
\end{array}\right), \qquad
F^{33}=\left(
\begin{array}{c c c}0&0&0\\
 0& 1&0\\
0&0&-1
\end{array}\right),
\eea
where ${\rm id}_3$ is the identity operator. The quasi-local charge (\ref{final_result}) can be rewritten as
\bea
\fl \mathcal{Q}(\lambda)= \sum_{\left\{a_j\right\},\left\{b_j\right\}}\left(\tr_{\mathcal{A}_m}\left\{\frac{{\rm d}}{{\rm d}v} \left(M_{a_N b_N}(\lambda)\ldots M_{a_1 b_1}(\lambda)  \right)\Big|_{v=0} \right\}\prod_{j=1}^{N}F_{j}^{a_jb_j}\right)
\nonumber \\
\hspace{5cm}- \tr_{\mathcal{A}_m} \left\{ \frac{{\rm d}}{{\rm d}v}\left(M_{11}(\lambda) \right)^N\Big|_{v=0} \right\}\,.
\label{new_result}
\eea
Here, we have defined
\bea
M_{\alpha\beta}(\lambda)=\widetilde{L}_{\alpha\beta}(\lambda),\quad {\rm if\  }\alpha\neq \beta\\
M_{11}(\lambda)=\frac{1}{3}\left(\widetilde{L}_{11}(\lambda)+\widetilde{L}_{22}(\lambda)+\widetilde{L}_{33}(\lambda)\right),\\
M_{22}(\lambda)=\frac{2}{3}\widetilde{L}_{11}(\lambda)-\frac{1}{3}\widetilde{L}_{22}(\lambda)-\frac{1}{3}\widetilde{L}_{33}(\lambda),\\
M_{33}(\lambda)=-\frac{1}{3}\widetilde{L}_{11}(\lambda)+\frac{2}{3}\widetilde{L}_{22}(\lambda)-\frac{1}{3}\widetilde{L}_{33}(\lambda).
\eea
Next, note that
\bea
M_{11}(\lambda)\Big|_{v=0}|0\rangle =|0\rangle,\label{aeq:1}\\
M_{\alpha\beta}(\lambda)\Big|_{v=0}|0\rangle =0 \quad {\rm otherwise.}\label{aeq:2}
\eea
Using Eqs. (\ref{new_result}), (\ref{aeq:1}), (\ref{aeq:2}), $\mathcal{Q}(\lambda)$ can be rewritten as
\be
\mathcal{Q}(\lambda)=\sum_{r=1}^{N}\sum_{x=0}^{N-1}\mathcal{P}^{x}\left(q_{r}(\lambda)\otimes{\rm id}_{N-r}\right)+\sum_{x=0}^{N-1}\mathcal{P}^{x}\left(R_N(\lambda)\right),
\label{eq:densities}
\ee
where ${\rm id}_{N-r}$ is the identity operator over the sites $r+1,\ldots, N$ and $\mathcal{P}^{x}$ is the translation operator (\ref{def:translation}). The density $q_r(\lambda)$, which is supported over the sites $1, \ldots ,r$, is given for $r\geq 2$ by
\bea
\fl q_r(\lambda)&=&\sum_{\left\{a_j\right\},\left\{b_j\right\}}\langle0|\mathcal{M}_{-}(\lambda) M_{a_{r-1} b_{r-1}}(\lambda)\ldots M_{a_2 b_2}(\lambda) \mathcal{M}'_{+}(\lambda)|0\rangle\prod_{j=2}^{r-1}F_{j}^{a_jb_j}\nonumber\\
\fl &=&\sum_{\left\{a_j\right\},\left\{b_j\right\}}\langle0|\mathcal{M}_{-}(\lambda) \widetilde{L}_{a_{r-1} b_{r-1}}(\lambda)\ldots \widetilde{L}_{a_2 b_2}(\lambda)  \mathcal{M}'_{+}(\lambda)|0\rangle\prod_{j=2}^{r-1}E_{j}^{a_jb_j},
\label{eq:representation}
\eea
where
\bea
\mathcal{M}_{-}(\lambda)&=&M_{12}(\lambda)F_r^{12}+M_{13}(\lambda)F_r^{13}+M_{23}(\lambda)F_r^{23},\\
\mathcal{M}'_{+}(\lambda)&=&\frac{{\rm d}}{{\rm d} v}\left\{M_{21}(\lambda)F_1^{21}+M_{31}(\lambda)F_1^{31}+M_{32}(\lambda)F_1^{32}\right\}\Big|_{v=0}.
\eea
Instead, for $r=1$, we have
\bea
 q_1(\lambda)&=&\langle0|\frac{{\rm d}}{{\rm d} v}\left(M_{22}(\lambda)F_1^{22}+M_{33}(\lambda)F_1^{33}\right)\Big|_{v=0}|0\rangle.
\eea
Analogously, the operator $R_N(\lambda)$ in Eq. (\ref{eq:densities}) is defined as
\bea
\fl R_N(\lambda)= \sum_{\left\{a_j\right\},\left\{b_j\right\}}\left(\sum_{k=1}^{m-1}\langle k|\frac{{\rm d}}{{\rm d}v} \left(M_{a_N b_N}(\lambda)\ldots M_{a_1 b_1}(\lambda)  \right)\Big|_{v=0} |k\rangle\prod_{j=1}^{N}F_{j}^{a_jb_j}\right)
\nonumber \\
\hspace{5cm}- \sum_{k=1}^{m-1}\langle k| \frac{{\rm d}}{{\rm d}v}\left(M_{11}(\lambda) \right)^N\Big|_{v=0}|k \rangle.
\eea
From the representation (\ref{eq:representation}) it is now possible to see that $||q_r(\lambda)||_{\rm HS}^2$ is exponentially vanishing with $r$, as required from Eq.~(\ref{def:densities}). This follows straightforwardly from the properties (\ref{prop_2}), (\ref{prop_3}) of the matrix $\mathcal{L}_{\mathcal{A}\mathcal{A}}$ discussed in section~\ref{sec:main_result}. Similarly, it can be seen that $||R_{N}(\lambda)||^{2}_{\rm HS}$ is exponentially vanishing with the size $N$ of the system, and it is thus a remainder that vanishes in the large $N$ limit. This remainder is analogous to that appearing in the quasi-local charges derived in the spin-$1/2$ case in Ref. \cite{prosen}. Putting everything together, we obtain that the charges (\ref{final_result}) are indeed quasi-local according to the definition encoded in Eqs.~(\ref{def_quasi_locality}), (\ref{def:densities}).

\end{document}